\documentclass[10pt]{article}

\usepackage{lmodern}

\usepackage[small,compact,center]{titlesec}

\usepackage{setspace}
\usepackage{longtable}

\usepackage[font={small,it}]{caption}

\usepackage[letterpaper,pass,includehead,includefoot,includeheadfoot,showframe=false]{geometry}
\usepackage{fullpage}

\usepackage{multicol}

\usepackage{amsmath,amsfonts,amssymb,mathtools}
\allowdisplaybreaks[1]	

\usepackage{color}

\usepackage{enumitem}\setlist{nosep}
\usepackage[colorlinks = true,
            linkcolor = blue,
            urlcolor  = blue,
            citecolor = blue,
            anchorcolor = blue]{hyperref}

\usepackage{cleveref}
\usepackage{blkarray}

\usepackage{graphicx}

\usepackage{subcaption}
\usepackage[font={small,it}]{caption}

\usepackage{fancyhdr}

\usepackage{algorithm}
\usepackage{algpseudocode}

\usepackage{authblk}

\title{Quantitative assessment of biological dynamics with aggregate data}

\author[1]{Stephen McCoy}
\author[1]{Daniel McBride}
\author[2]{D.~Katie McCullough}
\author[2]{Benjamin C.~Calfee}
\author[2]{Erik Zinser}
\author[2]{David Talmy}
\author[1,*]{Ioannis Sgouralis}

\affil[1]{Department of Mathematics, University of Tennessee Knoxville, Knoxville, TN}
\affil[2]{Department of Microbiology, University of Tennessee Knoxville, Knoxville, TN}

\affil[*]{Corresponding author, isgoural@utk.edu}

\date{}

\def\R{\mathbb{R}}
\def\half{\frac{1}2}
\def\grad{\nabla}
\def\HM{\mathcal{M}}
\def\l{\lambda}

\def\Gam{\text{Gamma}}
\def\LogN{\text{LogNormal}}

\def\d{\delta}
\def\g{\gamma}
\def\k{\kappa}
\def\th{\theta}

\def\HN{\mathcal{N}}

\begin{document}

\maketitle
\vfill

\begin{abstract}
We develop and apply a learning framework for parameter estimation in initial value problems that are assessed only indirectly via aggregate data such as sample means and/or standard deviations. Our comprehensive framework follows Bayesian principles and consists of specialized Markov chain Monte Carlo computational schemes that rely on modified Hamiltonian Monte Carlo to align with constraints induced by summary statistics and a novel elliptical slice sampler adapted to the parameters of biological models. We benchmark our methods with synthetic data on microbial growth in batch culture and test them with real growth curve data from laboratory replication experiments on \textit{Prochlorococcus} microbes. The results indicate that our learning framework can utilize experimental or historical data and lead to robust parameter estimation and data assimilation in ODE models that outperform least-squares fitting.

\textbf{Keywords:} Batch culture, growth curve, \textit{Prochlorococcus}, dynamical systems, Hamiltonian Monte Carlo, statistical learning
\end{abstract}

\vfill

\newpage

\section{Introduction}

The practice of acquiring large sets of individual data points and combining them to obtain diverse summary statistics, which we refer to as \emph{data aggregation,} is a commonly used technique in multiple domains, including economics, policy making, social sciences, health care and biological and ecological research \cite{jim2020,venkataramani2020,markham2023}. In modern science and engineering, the interpretation of aggregate data is often preferable to the interpretation of raw data, and this practice is represented in all high-level data analysis contexts.

Specifically, in the life and biological sciences aggregate data is often generated in replication experiments where the experiment's actual raw measurements are used to generate summary statistics, such as averages and standard errors, that are maintained, curated, and made openly available while the original raw measurements are either discarded or access to them is kept restricted. This practice has the obvious advantages of conserving storage space when processing large data sets such as time-lapse and image data \cite{lee2017}, obscuring the actual source of the data for security and privacy or competitive purposes \cite{wilson2021,haibe2020}, reducing noise \cite{xu2020,diaz2017,sgouralis2016transfer,sgouralis2015mathematical,sgouralis2013control}, and aiding comparison between different experiments or protocols that are probing the same system but employing different modalities \cite{xia2022,sgouralis2017introduction,sgouralis2013control,sgouralis2014theoretical}. However, such practices can cause problems; for example, many Covid-19 data related to maternal and neonatal outcomes are restricted to only summary statistics, leaving clinicians and patients to operate on partial information \cite{smith2020}.

An additional challenge, perhaps more significant, of working with aggregate data is the loss of information that occurs when it is generated \cite{orcutt1968} and also the loss of reference to the underlying specific biological processes \cite{ronan2016,raue2013lessons}. For example, similar to averaging, the integration of signals that occurs during the acquisition of single-molecule fluorescence data obscures fast dynamics without allowing for the precise estimation of detailed kinetics \cite{kilic2021extraction,kilic2021residence,kilic2021continuous,kilic2021generalizing,mattamira2025bayesian}. In addition, biological specimens often exhibit heterogeneity, and aggregating data without accounting for heterogeneity may produce skewed results \cite{sgouralis2016bladder,sgouralis2017renal}. Accurate quantitative analysis requires careful experimental design and data processing to avoid masking subtle differences between individual specimens. Under these conditions, reproducibility is particularly challenging, as variability in biological systems can make replicated findings difficult due to small differences that remain uncharacterized prior to data acquisition.

Data aggregation also poses serious challenges with rigorous data analysis or assimilation techniques \cite{law2015}. Specifically, within data assimilation, the goal is to develop mathematical frameworks that process the information available in empirical form and obtain quantitative predictions \cite{williamson2002}. However, predictions following the assimilation of aggregated data are limited by lost information or distortions caused by aggregation \cite{heesche2022implications,clark1976effects}. In order to counteract such artifacts, elaborate frameworks that combine domain knowledge and physical constraints in the form of specialized models are often required. Such approaches naturally lend themselves to Bayesian assimilation methodologies \cite{reich2015probabilistic}.

Bayesian data assimilation methods have seen an increase in popularity, particularly in parameter estimation applications for dynamical problems modeled with \emph{initial value problems} (IVP) \cite{huang2020bayesian,hinson2023,linden2022}. A major advantage is that they offer uncertainty quantification, as the standard practice is to learn an entire distribution of plausible values for every variable of interest rather than point estimators. Another advantage is that they can restrict the parameters under estimation to only meaningful values; for instance, a rate parameter or an initial population can be assigned a prior distribution with only positive support, or prior distributions can be defined on only the intervals that remain meaningful in the spatiotemporal scales of the underlying problem. Furthermore, they apply equally to both identifiable and nonidentifiable cases, the distinction of which is a common modeling challenge in mathematical biology \cite{maiwald2016driving,wieland2021structural}. 
Finally, they offer modeling flexibility and realism that comes from explicitly accounting for different sources of noise and the complexity characteristic of the systems being studied \cite{presse2023}.

Nevertheless, Bayesian data assimilation is not without drawbacks, especially for IVPs \cite{rodriguez2006hybrid,schober2014probabilistic}. To start, one needs to construct and characterize, often via intensive computational sampling procedures, the relevant posterior distribution. In particular, in the parameter estimation problem of models with ordinary differential equations (ODE), this involves numerically integrating the ODE for each new configuration of parameters for each posterior sample \cite{murphy2024implementing,roda2020bayesian,almutiry2021continuous}. Given that the required posterior sample sizes are typically large, sampling strategies become computationally costly and require efficient computational algorithms that are a topic of ongoing research. In addition, biological dynamics are typically contaminated with \emph{multiplicative noise} \cite{campbell1995,campbell1995lognormal,hinson2023}, which poses additional challenges to common algorithms that assume additive noise \cite{tronarp2019probabilistic,gabor2015robust}, such as sudden divergence of Kalman filters \cite{presse2023,briers2010smoothing} or degeneracy of particle filters \cite{djuric2003particle}. Furthermore, the predictions provided in Bayesian data assimilation depends critically on the dynamical model used, its fidelity to the modeled system, and the quality of the data supplied. 

A particular challenge when training a dynamical model with aggregated data is that, to properly model the data points that give rise to the supplied summary statistics, we have to mathematically reproduce the collapsing of unavailable raw measurements down to the available values. In the statistical representation of the resulting model, this translates into fitting the model parameters with \emph{singular distributions,} that is, probability distributions supported only on subspaces of lower dimension than the model's full parameter space. Parameter estimation is a routine topic in the Bayesian literature focused on simple models with nonsingular distributions \cite{sivia2006data,presse2023}; however, parameter estimation with singular distributions under complex statistical models necessary to tackle real-world scenarios remains an open challenge.

In this study, we develop a novel comprehensive statistical learning framework that addresses these challenges. Our framework allows for parameter estimation in IVPs of ODE models and also allows modeling of the latent raw measurements that gave rise to the summary statistics forming our dataset. To perform parameter estimation on our statistical model, we apply computational methods that can sample from distributions restricted to particular subsets of the parameter space determined by data aggregation. To this end, we develop a novel extension of the Hamiltonian Monte Carlo (HMC) sampling algorithm that allows for navigating highly dimensional parameter spaces while accounting for parameter constraints. In addition, we develop a novel slice sampling scheme that allows for the IVP parameter's positive support to be exploited efficiently by means of the elliptical slice sampling (ESS) algorithm.

The rest of this study is structured as follows. In \textsc{Methods,} we begin by describing the data format under consideration, the IVP we aim to address, and the Bayesian framework that we use for its parameter estimation. First, we present our framework in general form, followed by its application to the analysis of growth curve data. In \textsc{Results,} we demonstrate how our framework performs and compares it with standard estimation approaches, such as least squares fitting, with \textit{in silico} and \textit{in vivo} data of microbial growth curves from batch culture laboratory experiments commonly conducted in microbiology research. Lastly, in \textsc{Discussion,} we provide an overview of our methods and elaborate on its perspectives for future applications.

\section{Methods}

In this section, we first present a general description of our framework that emphasizes modeling and computational aspects. Subsequently, we present a specialized application to microbial growth curve data acquired in replication experiments.

\subsection{Statistical learning}

\subsubsection{Learning scheme}
\label{sec:learning}

Our framework considers the analysis of aggregated data that we denote by $z_{1:N}^{1:M}$. Specifically, we denote individual data points with $z_n^m$ and use superscripts $m=1,\dots,M$ to refer to different summary statistics that may be available and subscripts $n=1,\dots,N$ to refer to assessments made at time $t_n$. For example, $z_2^1$ indicates our 1\textsuperscript{st} summary statistic obtained at the 2\textsuperscript{nd} time assessment.

Our data stem from collapsing batches of raw measurements made at the same time, which we represent by
\begin{align}
z_n^m&=G^m\left(y_n^{1:K}\right)
.
\label{eq:stats}
\end{align}
Here, $y_n^k$ denotes individual raw measurements, made at time $t_n$, and $k=1,\dots,K$ indexes the batch of measurements. The functions $G^m(\cdot)$ model the corresponding batch statistics. For instance, the common sample mean and standard deviation correspond to the functions
\begin{align}
G^1\left(y^{1:K}\right)
&=
\frac{1}{K}\sum_{k=1}^{K} y^{k}
,
\label{G1func}
    \\
G^2\left(y^{1:K}\right)
&=
\sqrt{\frac{1}{K-1}\sum_{k=1}^{K} \left( 
       y^k-\frac{1}{K}\sum_{k'=1}^{K} y^{k'}
\right)^2}
.
    \label{G2func}
\end{align}
In this study, we focus on the cases where \emph{only} $z_{1:N}^{1:M}$ are available, while $y_{1:N}^{1:K}$ are \emph{not.} For this reason, we explicitly model the missing measurements using a likelihood that represents biological or measurement noise. Our likelihood, which we cast in the form
\begin{align}
y_{n}^{k}|g,h
\sim
\mathbb{A}\left(
F\left(x^g(t_n)\right),h\right)
,
\label{eq:like}
\end{align}
is a probability distribution that models the generation of independent measurements. Here, $F(\cdot)$ is a problem-specific observation function that links measurements with the dynamical variables $x^g(\cdot)$ of an underlying ordinary differential equation (ODE) that models the dynamics of interest. Often, this is simply a projection that reduces the full dynamical state of the system of interest to a single component. The parameter $h$ allows for tunable noise characteristics, such as noise spread, which can better reflect the inherent statistics of the measurements.

To model the dynamics of our system, we consider a generic initial value problem (IVP) for an ODE of the from
\begin{align}
\frac{dx}{dt}&=H^g(t,x)
\label{eq:ODE}
.
\end{align}
The dynamics function $H^g(\cdot,\cdot)$ is also problem specific and depends on unknown parameters gathered in $g$. Together, with the appropriate initial conditions that may also depend on unknown parameters gathered in $g$, our IVP leads to a solution $x^g(t)$ which, at any time $t$, describes the dynamical state of the system of interest.

Our framework contains \emph{unknown measurements} $y_{1:N}^{1:K}$ whose statistical properties are fully described by the likelihood in \cref{eq:like} and also \emph{unknown parameters} $g$ and $h$. The latter is associated with the noise of the system, while the former is associated with the dynamics. Following the Bayesian paradigm \cite{presse2023}, we assign priors to them
\begin{align}
h&\sim\mathbb{B},
\label{eq:prior_g}
\\
g&\sim\mathbb{C},
\label{eq:prior_c}
\end{align}
which, in addition to completing our statistical framework, allow the specification of meaningful bounds for their values through their support.

\begin{figure}[tbp]
    \centering
\includegraphics[scale=1]{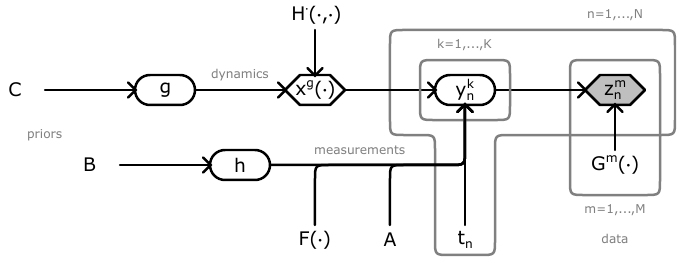}
    \caption{Graphical representation of the statistical learning framework in this study. Following the common convention \cite{presse2023, BN2006}, random quantities are shown with circles, deterministic quantities are shown with diamonds, and quantities with known values are shaded. In addition, arrows indicate dependencies among the various quantities of interest, plates indicate repetition over the marked index, and hyperparameters are left free.}
    \label{fig:frame}
\end{figure}

Given a data set $z_{1:N}^{1:M}$, assessment times $t_{1:N}$, statistics $G^{1:M}(\cdot)$, and batch size $K$,
our learning framework consists of problem-specific choices for the noise distribution $\mathbb{A}(\cdot,\cdot)$, noise parameter $h$, observable $F(\cdot)$, dynamical variables $g$, dynamics $H^\cdot(\cdot,\cdot)$, and prior distributions $\mathbb{B},\mathbb{C}$. A graphical summary is shown in \cref{fig:frame} and a complete statistical summary is given in \cref{sec:append_eqn}.

Once such choices are made, our framework leads to a posterior probability distribution which, via Bayes' rule
\cite{gelman1995,lee1989}, is formally characterized by a probability density that takes the form
\begin{align*}
\mathcal{P}\left(h,g,y_{1:N}^{1:K}|z_{1:N}^{1:M}\right)
&\propto
\mathcal{P}\left(z_{1:N}^{1:M}|y_{1:N}^{1:K}\right)
\mathcal{P}\left(y_{1:N}^{1:K}|g,h\right)
\mathcal{P}\left(h\right)
\mathcal{P}\left(g\right)
.
\end{align*}
According to our model's statistical representation, this density specializes to
\begin{align}
\mathcal{P}\left(h,g,y_{1:N}^{1:K}|z_{1:N}^{1:M}\right)
&\propto
\left[
\prod_{n=1}^N
\left(
\prod_{m=1}^M
\delta_{G^m\left(y_n^{1:K}\right)}\left(z_n^m\right)
\right)
\left(
\prod_{k=1}^K
A\left(y_n^k|g,h\right)
\right)
\right]
B\left(h\right)
C\left(g\right)
.
\label{eq:posterior}
\end{align}
Here, the factors that contain Dirac deltas $\delta_{G^m\left(\cdot\right)}\left(\cdot\right)$ arise due to \cref{eq:stats}, which dictates the precise agreement between our model measurements and the corresponding batch statistics, while $A(\cdot|\cdot,\cdot),B(\cdot),C(\cdot)$ are the probability functions associated with the distributions $\mathbb{A}(\cdot,\cdot),\mathbb{B},\mathbb{C}$ in \cref{eq:like,eq:prior_g,eq:prior_c}, respectively.

\subsubsection{Markov chain Monte Carlo}
\label{sec:MCMC}

Due to the IVP, which most often remains analytically intractable, we generally cannot obtain a closed-form expression to the posterior of \cref{eq:posterior}. For this reason, to characterize \cref{eq:posterior}, we develop a specialized Markov chain Monte Carlo (MCMC) sampling scheme that generates pseudorandom samples \cite{L2001,RC1999,metropolis1953} of the unknowns of the model $h,g,y_{1:N}^{1:K}$.

Due to the natural grouping of our unknowns, we use a \emph{Gibbs sampler} that iterates the following two steps:
\\
\textbullet~update \emph{parameters} by sampling from $\mathcal{P}\left(h,g|y_{1:N}^{1:K},z_{1:N}^{1:M}\right)$,
\\
\textbullet~update \emph{measurements} by sampling from $\mathcal{P}\left(h,y_{1:N}^{1:K}|g,z_{1:N}^{1:M}\right)$.
\\
To initialize our sampler, we directly generate $h$ and $g$ from the priors $\mathbb{B}$ and $\mathbb{C}$, respectively. Next, we generate $y_n^k\mid g,h,$ from $\mathbb{A}$, and then apply root-finding on the generated $y_{1:N}^{1:M}$ to ensure that the constraints of \cref{eq:stats} are satisfied. In all cases, we combine our sampler with appropriate numerical integrators, such as adaptive Runge-Kutta or other specialized schemes, to solve the IVP according to the specifics of the problem at hand \cite{quarteroni2006,stoer1980,atkinson1991,shampine1997}. 

In the special case where $\mathbb{A}$ and $\mathbb{B}$ are conditionally conjugate \cite{gelman1995}, our Gibbs updates can be implemented via ancestral sampling
\cite{BN2006,L2001,gelman1995,presse2023}
based on the respective factorizations
\begin{align*}
\mathcal{P}\left(h,g|y_{1:N}^{1:K},z_{1:N}^{1:M}\right)
&=
\mathcal{P}\left(h|g,y_{1:N}^{1:K}\right)
\mathcal{P}\left(g|y_{1:N}^{1:K}\right),
\\
\mathcal{P}\left(h,y_{1:N}^{1:K}|g,z_{1:N}^{1:M}\right)
&=
\mathcal{P}\left(h|g,y_{1:N}^{1:K}\right)
\mathcal{P}\left(y_{1:N}^{1:K}|g,z_{1:N}^{1:M}\right),
\end{align*}
both of which allow direct generation of $h$ via $\mathcal{P}\left(h|g,y_{1:N}^{1:K}\right)$. For both the parameters and the measurements, our updates are derived from specialized samplers adapted to $\mathcal{P}\left(g|y_{1:N}^{1:K}\right)$ and $\mathcal{P}\left(y_{1:N}^{1:K}|g,z_{1:N}^{1:M}\right)$ as described below. 

\paragraph{mESS for parameter sampling}

The first Gibbs update requires the generation of $g$ given measurements $y^{1:K}_{1:N}$ from $\mathcal{P}\left(g|y_{1:N}^{1:K}\right)$, which we achieve using a novel \emph{multiplicative elliptical slice sampler} (mESS), which naturally aligns with distributions restricted to positive support, as often found in dynamical systems of biological processes. See \cref{sec:append_MESS} for a detailed explanation of the sampler. Our approach retains the core benefits of elliptical slice sampling, such as parameter updates without tuning and efficient exploration of complex posterior landscapes, while offering a targeted enhancement for distributions with strictly positive values and non-Gaussian characteristics \cite{presse2023,murray2010elliptical}. 

\paragraph{cHMC for measurement sampling}

In our second Gibbs update, we generate samples of $y_{1:N}^{1:K}$ given known parameters $g$ from
$\mathcal{P}\left(y_{1:N}^{1:K}|g,z_{1:N}^{1:M}\right)$ that satisfy the constraints $z_n^m = G^m(y_n^{1:K})$ of \cref{eq:stats}. For this, we apply a novel \emph{constrained Hamiltonian Monte Carlo} (cHMC) sampler specialized for handling the constraints. Our method is fully detailed in \cref{sec:append_cHMC}. As we explain in \cref{sec:append_HMC}, standard HMC is suitable for sampling smooth high-dimensional distributions with full support \cite{L2001, GJM2011, betancourt2017conceptual}; however, satisfying the constraints requires modifications as described in \cref{sec:append_cHMC}. Our novel cHMC sampler takes advantage of the RATTLE numerical integrator \cite{A1983,B2012,HLW2006} in the HMC integration loop to ensure that the generated $y_{1:N}^{1:K}$ satisfy the constraints while remaining statistically correct, i.e.~ensuring that our MCMC chain converges to the target distribution of \cref{eq:posterior}. Our method maintains HMC's efficient sampling of high-dimensional distributions \cite{L2001, GJM2011, betancourt2017conceptual} arising due to multiple time points (i.e.~$N\gg1$) and large batch-sizes (i.e.~$K\gg1$) per time point, while permitting navigation on the support of singular distributions arising due to the constraints.

\subsection{Application to Prochlorococcus growth curve data}
\label{sec:meth_pro}

In this section, we specialize our statistical learning framework in a case study of interest in microbiology and marine biology. In our study, the aggregated data stem from batch growth experiments of the marine cyanobacteria \textit{Prochlorococcus} (\textit{Pro}). \textit{Pro} is the most abundant photosynthetic phytoplankton in the ocean \cite{partensky1999proc} and is studied  \textit{in situ} for genetic and physical connections to their bio-geographical significance \cite{flombaum2013present,berube2015physiology}. These photosynthetic organisms play a vital role in the regulation of ocean food chains and climate \cite{biller2015,partensky1999proc}. Laboratory growth experiments take the form of replicate time series data, where multiple sets of triplicate test tubes of \textit{Pro} are monitored for cell density as they grow in batch culture. Data aggregation is applied to the runs to produce sample averages and standard deviations.

Our growth curves depend on two pivotal quantities: the maximum growth rate, typically reported in units of $1/\text{\rm days}$, and the nutrient affinity, typically reported in units $\text{\rm ml} / (\text{\rm cells}\cdot\text{\rm days})$, of the \textit{Pro} cells which we denote with $m$ and $a$, respectively. These parameters, as well as the initial nutrient and \textit{Pro} cell densities of batch culture experiments, which we denote with $Q$ and $P$, determine the overall dynamics of the cell and nutrient density.

To model the dynamics in \cref{eq:ODE} we introduce an IVP that consists of
\begin{align}
    \frac{dq}{dt} &= -\frac{q}{q+m/a}mp & q(t_{0}) &= Q \label{qODE} \\
    \frac{dp}{dt} &= +\frac{q}{q+m/a}mp & p(t_{0}) &= P \label{pODE}
\end{align}
for a fixed initial time \( t_{0} \) that coincides with the onset of the experiment. Our dynamical state $x=(q,p)$ consists of the density of nutrients $q$ and the density of cells $p$, both reported in $\text{\rm cells/ml}$. Our \cref{qODE,pODE} depend on the unknown parameters $g = (Q,P,m,a)$. Given $g$, we denote the solution of our IVP with $x^g(t)=(q^g(t),p^g(t))$.

In a typical experiment that monitors the growth curve, the measurements probe only the cell density. Accordingly, our measurement function in \cref{eq:like} reduces to a projection
\begin{align}
F(q,p)=p
.
\end{align}
Theoretical and empirical studies on microbial growth \cite{campbell1995,campbell1995lognormal,hinson2023} indicate the presence of multiplicative noise in the measurements. Accordingly, for the likelihood $\mathbb{A}(\cdot,\cdot)$ we choose the LogNormal distribution
\begin{equation}
    y_{n}^{k}|g,h
    \sim
    \LogN\left(
    p^g(t_n),h\right)
    ,
    \label{ydisr}
\end{equation}
with an unknown scale parameter $h$. For the definition, see \cref{sec:dist_def}. Following \cref{sec:learning}, we denote by $y_n^k$ the $k^\text{th}$ measurement made at time $t_n$ during the experiment and assume the statistics in \cref{G1func,G2func} to form the reported batch mean $z^1_n$ and standard deviation $z^2_n$ of each time point.

Finally, for \cref{eq:prior_g}, we apply a prior on the parameters $g = (Q,P,m,a)$ of our IVP that allows only for strictly positive values as defined by 
\begin{align}
    Q &\sim \Gam\left(\phi_Q,\psi_Q/\phi_Q\right), \label{eq:prior_Q}\\
    P &\sim \Gam\left(\phi_P,\psi_P/\phi_P\right), \label{eq:prior_P}\\
    m &\sim \Gam\left(\phi_m,\psi_m/\phi_m\right), \label{eq:prior_m}\\
    a &\sim \Gam\left(\phi_a,\psi_a/\phi_a\right), \label{eq:prior_a}
\end{align}
and, for \cref{eq:prior_c}, we also apply a Gamma prior 
\begin{equation}
    h \sim \Gam
    \left(
    \phi_h,
    \psi_h/\phi_h
    \right),
\end{equation}
which is conditionally conjugate to the likelihood in \cref{ydisr}. Here, we use the parameterization of the Gamma distribution, shown in \cref{sec:dist_def}, which allows the specification of shape and expectation hyperparameters via $\phi_Q,\phi_P,\phi_m,\phi_a,\phi_h$ and $\psi_Q,\psi_P,\psi_m,\psi_a,\psi_h$, respectively.

Our framework leads to the derivation of a formal posterior density
\begin{equation}
\mathcal{P}\left(Q,P,m,a,y_{1:N}^{1:K}|z_{1:N}^{1:M}\right)
\propto
\mathcal{P}(Q)\mathcal{P}(P)
\mathcal{P}(m)\mathcal{P}(a)
\int_0^\infty dh\,
\mathcal{P}\left(z_{1:N}^{1:M},h|Q,P,m,a,y_{1:N}^{1:K}\right)
.
\label{eq:marg_posterior}
\end{equation}
Here, we marginalize the noise parameter $h$ since its value is of little biological interest. By this marginalization, we avoid the ancestral sampling steps in our MCMC scheme of \cref{sec:MCMC}, but still apply mESS to update $Q,P,m,a$ and cHMC to update $y_{1:N}^{1:M}$ as previously described.

\subsection{Data acquisition}
\label{sec:data_aquisition}

The \textit{in vivo} data shown in \textsc{Results} are obtained with the following methodology. Axenic cultures of \textit{Prochlorococcus} strain MIT9215 cyanobacteria were maintained in an artificial seawater medium, AMP-MN \cite{calfee2022prochlorococcus}, a
derivative of AMP-A medium \cite{morris2011dependence,moore2007culturing} without any nitrogen amendment. Purity tests to determine the axenicity of cyanobacteria stocks and experimental cultures were performed routinely as previously described in \cite{morris2008facilitation}. All experiments were carried out at 24$^\circ$C in Percival I36VLX incubators (Percival, Boone, IA) with modified controllers that allowed a gradual increase and decrease of cool white light to simulate sunrise and sunset with a peak midday light intensity of 150~$\mu$mol quanta m$^{-2}$s$^{-1}$ on a 14~hr:10~hr light:dark cycle \cite{zinser2009choreography}. The abundance of cyanobacteria was
quantified by flow cytometry using a Guava EasyCyte 8HT flow cytometer (Millipore, Burlington, MA) with populations of \textit{Prochlorococcus} determined by their red fluorescence \cite{morris2008facilitation,cavender1998dual}. Raw measurements were obtained in batches of size $K=24$ for a total of $N=9$ time points spread over the duration of the experiment.

\section{Results}

In this section, we show how our framework performs on estimating the parameters of IVPs. To demonstrate its effectiveness in revealing the correct parameter values, we first validate our model on synthetic data, mimicking the characteristics of real data, that are generated with prescribed parameter values. Our \textit{in silico} experiments are conducted by simulating the model of \cref{sec:meth_pro}. Subsequently, we demonstrate that our methods maintain their performance on real laboratory data. Our \textit{in vivo} experiments are conducted as described in \cref{sec:data_aquisition}. We also compare against naive parameter estimation methods based on least-squares fitting.

\subsection{In silico growth curve data}
\label{sec:in_silico}

\begin{figure}[tbp]
    \centering
    \includegraphics[width=\textwidth]{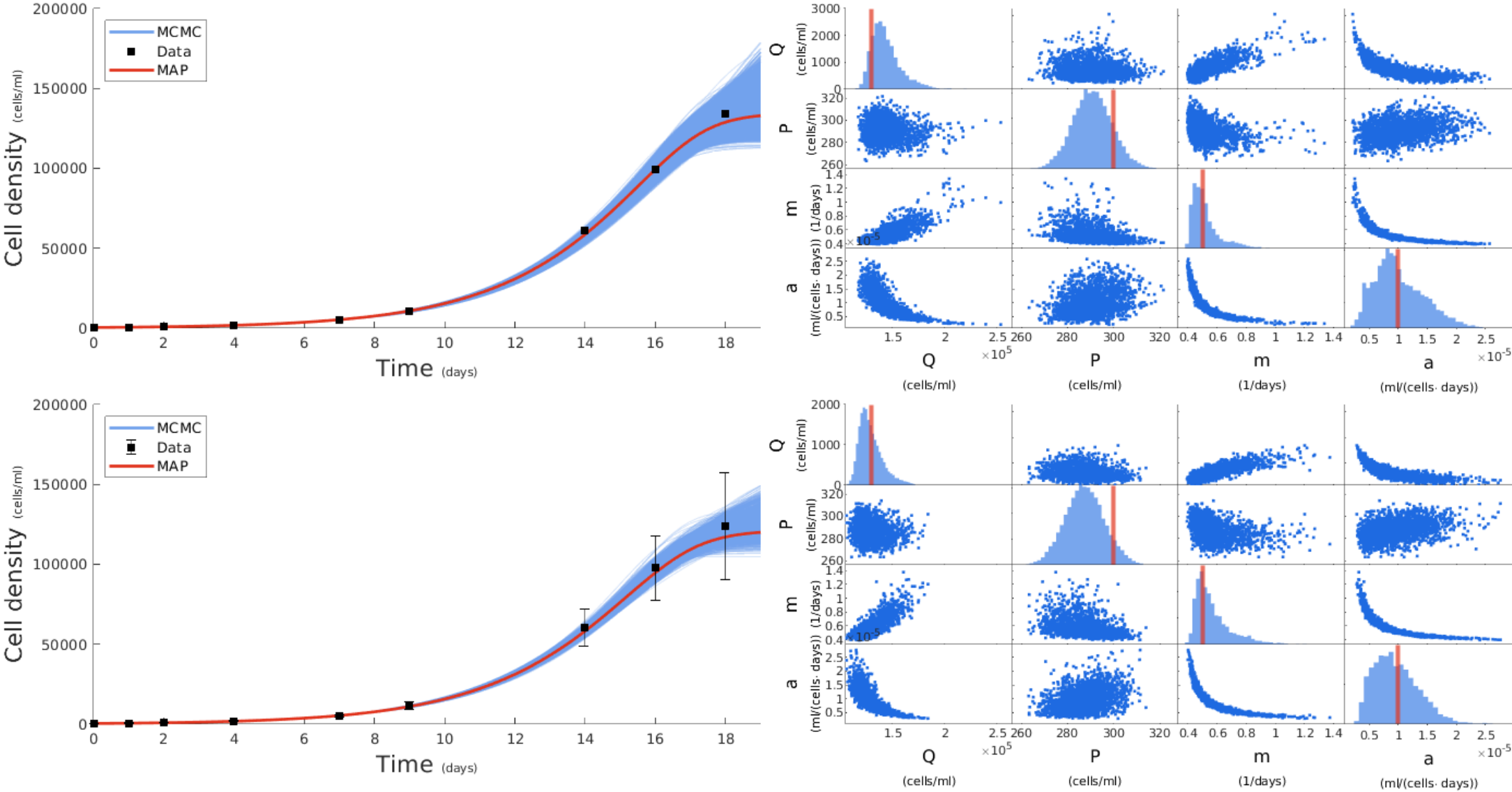}
    \caption{Fitting \emph{in silico} data with our statistical learning framework. The first row shows our results for a mean only constraint, and the second row shows our results for a mean and standard deviation constraint. Left: We use our MCMC parameter values to generate IVP solutions which are plotted as the blue curves, we then use the MAP estimate to generate the red curve and compare to the data points shown in black. Right: in the off-diagonal panels, we show MCMC scatter plots of pairwise parameter comparisons and, in the diagonal, we show MCMC histograms of each parameter and overlay the ground truth values with vertical red lines. Our results recover well the parameter truth parameters values and producing well-fitting IVP solutions.}
    \label{fig:ODE_param_est}
\end{figure}

To generate the synthetic data, we employ the IVP in \cref{pODE,qODE} modeling Pro growth with the values of the ground truth parameter $(Q,P,m,a) = (130\,000,300,0.5,0.000\,01)$ which we chose in agreement with the \emph{in vivo} data of the next section. Then, we generate cell density measurements $y_{1:N}^{1:K}$ according to \cref{ydisr} and derive summary statistics $z_{1:N}^{1:M}$ calculated by \cref{G1func,G2func}. Our data are shown in \cref{fig:ODE_param_est} with the upper panels corresponding to a scenario in which only batch means are maintained and the lower panels to a scenario in which both batch means and standard deviations are maintained after aggregation. We then employ our statistical learning framework to generate samples of the posterior distribution considering only sample means $\mathcal{P}(Q,P,m,a|z_{1:N}^1)$ or sample means and standard deviations $\mathcal{P}(Q,P,m,a|z_{1:N}^{1:2})$.

To allow for comparison with ground truth, for each case, we approximate the maximum a posterior estimate (MAP) of our parameters by selecting the MCMC sample with the highest posterior probability density \cite{van2011} which can be evaluated through \cref{eq:marg_posterior}. These estimates represent our framework's best choices for parameter values in each case and are indicated by solid lines in \cref{fig:ODE_param_est}. Their specific values are $(\hat Q,\hat P,\allowbreak\hat m,\hat a) \allowbreak = (134\,000,\allowbreak290,\allowbreak0.491,\allowbreak0.000\,0104)$ for the first case considering only batch means and $(\hat Q,\hat P,\allowbreak\hat m,\hat a) \allowbreak = (121\,000,\allowbreak291,\allowbreak0.509,\allowbreak0.000\,0107)$ for the second case considering both batch means and standard deviations. Both estimators are in good agreement with the ground truth.

For the two groups of panels in the left column of \cref{fig:ODE_param_est}, we also show a collection of sample trajectories. These are randomly selected MCMC samples that correspond to the solutions of \cref{qODE,pODE}. For both cases, we obtain a spread of trajectories around the MAP solution that indicates uncertainty coming from the noise in the measurements and missing information due to data aggregation. The uncertainty in the upper left panel is higher than that in the lower left panel, as indicated by the wider spread of the sampler trajectories around MAP. This is expected behavior because we assimilate the same information (i.e.~batch means) plus additional information (i.e.~batch standard deviations) in the second case.

The right panels in \cref{fig:ODE_param_est} show MCMC samples of the parameters. Specifically, along the diagonals, we show the MCMC approximations (histograms) of the marginal posteriors $\mathcal{P}(Q|z_{1:N}^{1:M}),\mathcal{P}(P|z_{1:N}^{1:M}),\allowbreak\mathcal{P}(m|z_{1:N}^{1:M}),\mathcal{P}(a|z_{1:N}^{1:M})$ along with the MAP estimates as vertical lines. Again, although there is general agreement with the ground truth, there is more uncertainty in the case of only the batch means than in the case of both batch means and standard deviations, as indicated by wider histograms. In the off-diagonal panels, we show MCMC approximations (scatter plots) of each pairing of the sample parameters. In both cases, our framework reveals preferences between combinations of parameters, indicating that, to comply with the supplied data, the trajectories of the underlying IVP require specific configurations of parameter values. For instance, the pair $(m,a)$ has the highest correlation among all pairs. This is expected behavior, as our dynamical model, see \cref{qODE,pODE}, depends \emph{only} on the ratio $m/a$ and not separately on $a$; therefore, the value of $a$ can be estimated only relative to the value of $m$, as seen.

\begin{figure}[tbp]
    \centering
    \begin{subfigure}[b]{\textwidth}
        \centering
        \includegraphics[width=\textwidth]{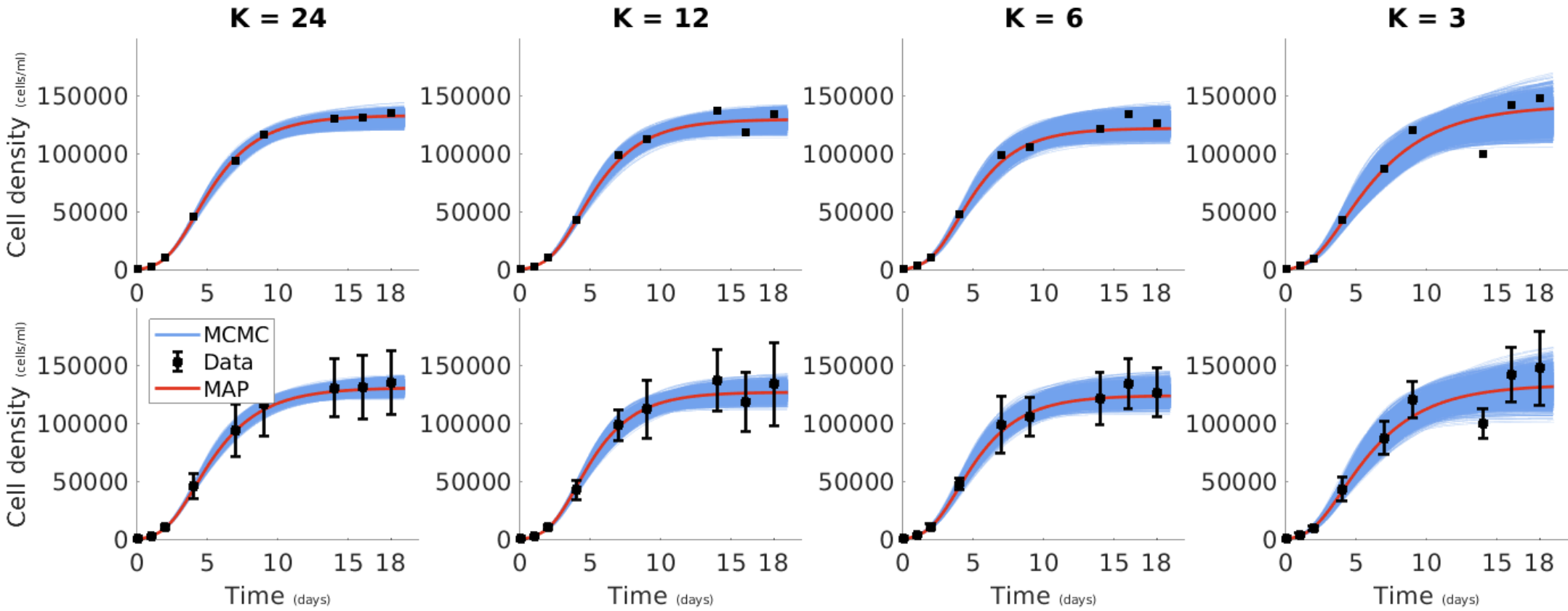}
        \caption{Data set size effect on IVP solutions}
    \end{subfigure}
    \hfill
    \begin{subfigure}[b]{\textwidth}
        \centering
        \includegraphics[width=\textwidth]{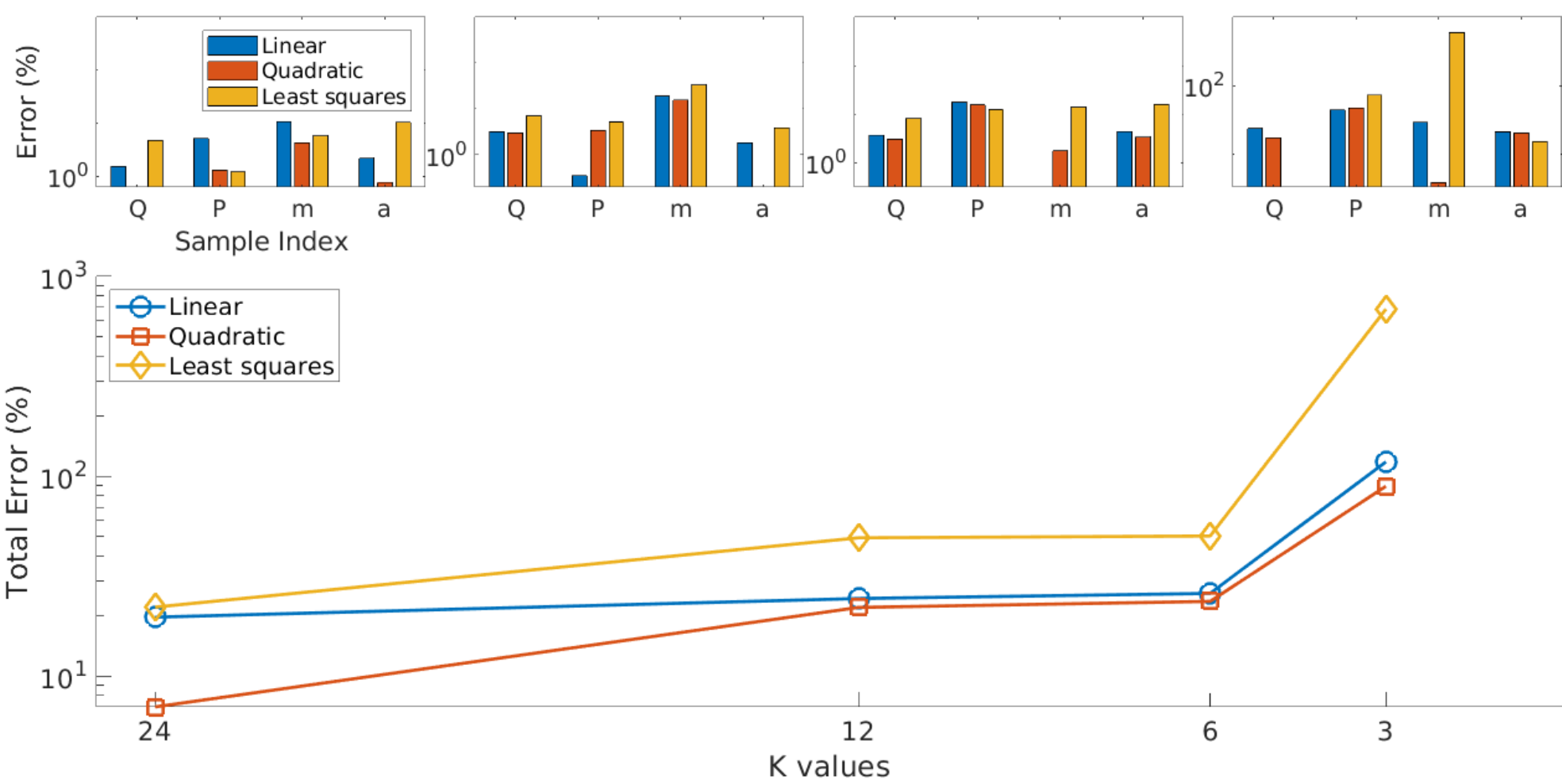}
        \caption{Data set size effect on error}
    \end{subfigure}
         \caption{Batch size robustness of our methods. We show the effect of reducing the batch size $K$ from $24$ to $3$ going from left to right across all figures.
         In the upper two rows, we show the solutions of our IVP \cref{qODE,pODE} for the constraints on the batch means (top) and the constraints on the batch means and standard deviations (bottom) as $K$ is decreased. In the bottom two rows, we show the percentage error between the mean only constraint MAP predictions (blue), mean and standard deviation constraint MAP prediction (red), and a LS prediction (yellow) for the parameter values $(Q,P,m,a)$ compared to their ground truth. This is shown in the upper panels as their individual errors in the form of a bar chart and in the bottom panel as a line plot of the total error (sum of individual errors) for each batch size. As we reduce the number of samples per time point down from a collection of experiments $K=24$, to one experiment $K=3$ we see LS breaks down whereas our framework copes better with the decrease in the signal to noise content of the resulting data. See \cref{tab:param_est} for the precise values used in the comparison.}
         \label{fig:LS comp}
\end{figure}

In addition to showing that our framework successfully recovers the values of the ground truth parameters with quantified uncertainty, in \cref{fig:LS comp} we demonstrate the performance of our methods when challenged with scenarios involving more demanding data. To this end, we apply our learning framework in the same two cases (only batch means and both batch means and standard deviations) now considering scenarios of \emph{decreasing} batch size $K=24,12,6,3$. In this way, we simulate a series of aggregate data generated in experiments of successively fewer raw measurements. Although fewer raw measurements result in increasingly noisier aggregate data, as seen in \cref{fig:LS comp} (upper two rows), our framework's estimates remain in good agreement with the ground truth. This indicates our framework's robustness to increased noise.

A head-to-head comparison with naive parameter estimation procedures mediated by least-square fitting (LS), see \cref{sec:append_LS}, indicates superior performance. In particular, in \cref{fig:LS comp} (bottom two rows) and \cref{tab:param_est} we quantify the percentage error in our MAP estimates and those resulting from LS for each parameter. For clarity, our error metrics are given by
\begin{equation*}
   \% \text{Error}
     = 
\left|
\frac{ X^{\text{\rm estimate}}}{
X^{\text{\rm ground}}
}
   - 1
\right|\times100\%
,
\end{equation*} 
where $X$ stands for any of $Q,P,m,a$. As seen, while the error for all three methods generally increases as the batch size $K$ decreases, and therefore the noise that persists in the aggregated data increases, LS consistently produces the least accurate estimates. In contrast, our framework consistently produces the most accurate ones. This indicates that, in addition to recovering the ground truth more accurately, our learning framework is also more robust when faced with excessively noisy data than existing practices.

Furthermore, a comparison between the batch means only scenario and the scenario incorporating both batch means and standard deviations, afforded by our framework, indicates that the best estimates are consistently obtained with the latter. Again, this is not surprising, since the latter scenario assimilates the largest amount of data.

\begin{table}[tbp]
\centering
\resizebox{\textwidth}{!}{ 
\begin{tabular}{|c|c|c||c|c||c|c||c|c|}
 \hline
 &       & Ground & \multicolumn{2}{|c||}{Only batch means} & \multicolumn{2}{|c||}{Both batch means and} & \multicolumn{2}{|c|}{Least squares} \\
 & Units & truth & \multicolumn{2}{|c||}{} & \multicolumn{2}{|c||}{standard deviations} & \multicolumn{2}{|c|}{} \\
 \hline
 $K=24$ & & & Estimate & $\%$ Error & Estimate & $\%$ Error & Estimate & $\%$ Error \\
 \hline
 $Q$ & $\text{cells}/\text{ml}$ & 130000 & 132000	&1.545	&131000	&0.630	&136000	&4.761

\\ 
 $P$ & $\text{cells}/\text{ml}$ & 300 &284	&5.173	&296	&1.311	&304	&1.216
 \\ 
 $m$ & $1/\text{days}$ & 0.5 &0.446	&10.828	&0.478	&4.300&	0.529	&5.811

 \\ 
 $a$ & $\text{ml} /(\text{cells} \cdot \text{days})$ & 0.00001 &.0000102&2.187	&.000010078&0.780&.00000896&10.372
 \\ 
 \hline
 $K=12$ & & & Estimate & $\%$ Error & Estimate & $\%$ Error & Estimate & $\%$ Error \\
 \hline
 $Q$ & $\text{cells}/\text{ml}$ & 130000 & 134000	&3.196	&134000	&2.985	&139000	&7.069
 \\ 
 $P$ & $\text{cells}/\text{ml}$ & 300  &301	&0.350	&310	&3.428	&316&	5.202
 \\ 
 $m$ & $1/\text{days}$ & 0.5 & 0.405	&19.074&0.423&15.462&0.335&33.054
 \\ 
 $a$ & $\text{ml} /(\text{cells} \cdot \text{days})$ & 0.00001 & .0000101&1.838	&.0000100&0.198&.0000103&3.898
 \\ 
 \hline
 $K=6$ & & & Estimate & $\%$ Error & Estimate & $\%$ Error & Estimate & $\%$ Error \\
 \hline
 $Q$ & $\text{cells}/\text{ml}$ & 130000 &135000&3.699&134000&3.041	&119000	&8.119
 \\ 
 $P$ & $\text{cells}/\text{ml}$& 300 & 353&17.526&347&15.541&263&12.440
 \\ 
 $m$ & $1/\text{days}$ & 0.5& 0.502&0.313&0.491&1.750&0.569&13.892
 \\ 
 $a$ & $\text{ml} /(\text{cells} \cdot \text{days})$ & 0.00001 &.00000955&4.406&.00000966&3.333&.0000115&15.797
 \\ 
 \hline
 $K=3$ & & & Estimate & $\%$ Error & Estimate & $\%$ Error & Estimate & $\%$ Error \\
 \hline
 $Q$ & $\text{cells}/\text{ml}$ & 130000 & 161000&23.737&153000&17.415&126000	&3.318
 \\ 
 $P$ & $\text{cells}/\text{ml}$ & 300 &432	&44.119&442&47.385&82&72.596
 \\ 
 $m$ & $1/\text{days}$ & 0.5 & 0.354&29.120&0.481&3.862&3.475&595.081
 \\ 
 $a$ & $\text{ml} /(\text{cells} \cdot \text{days})$ & 0.00001 & .00000788&21.138&.00000795&20.484&.0000115&15.305
\\
 \hline
\end{tabular}}
\caption{\label{tab:param_est} Quantitative validation against ground truth. The comparison shows that our model outperforms LS estimation, particularly as batch size $K$ size decreases.}
\end{table}

\subsection{In vivo growth curve data}
\label{sec:in_vitro}

\begin{figure}[tbp]
    \centering
    \includegraphics[width=\textwidth]{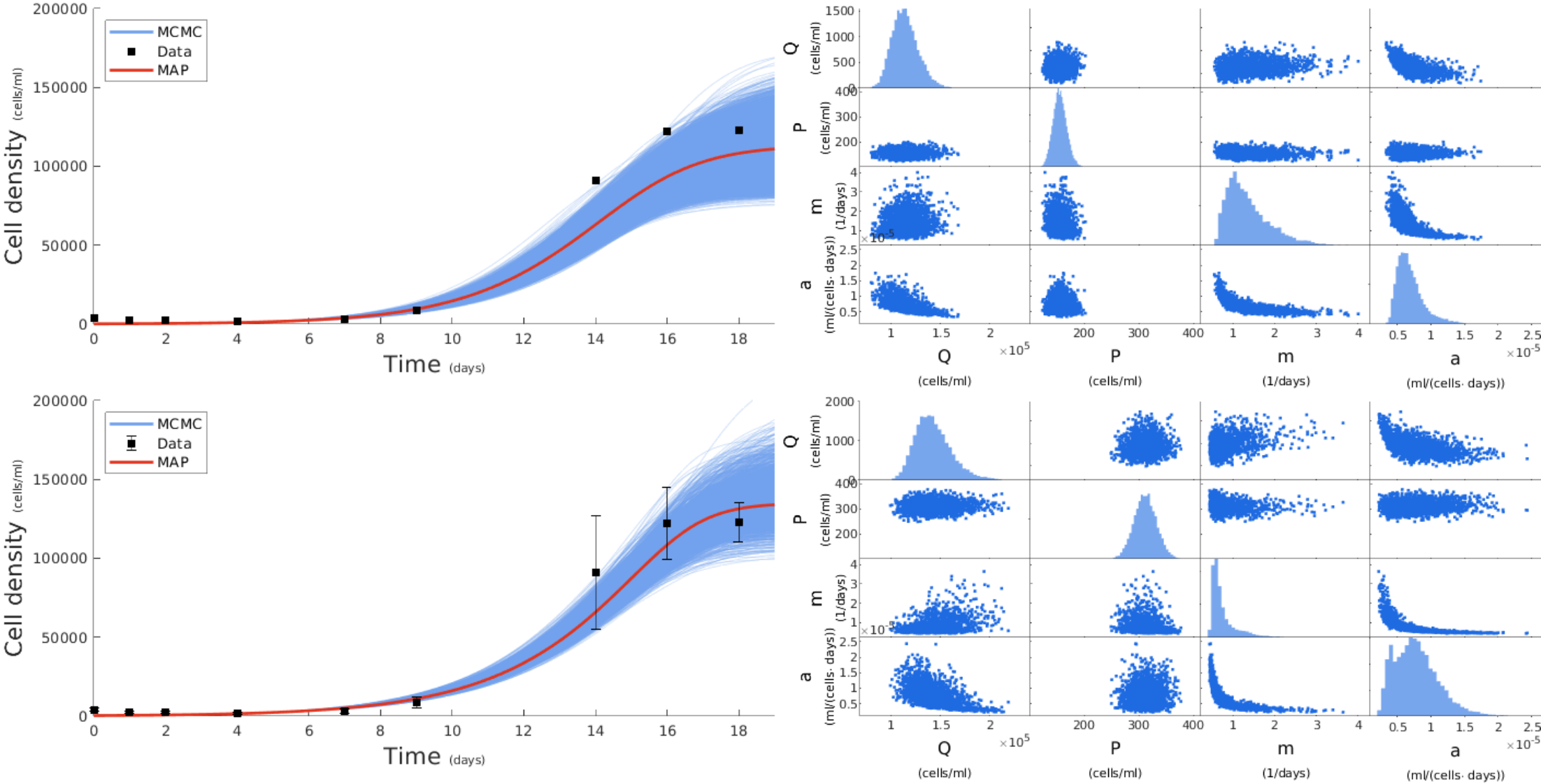}
    \caption{
    Fitting \emph{in vivo} data with our statistical learning framework. The first row shows our results for a mean only constraint, and the second row shows our results for a mean and standard deviation constraint. Left: We use our MCMC parameter values to generate IVP solutions which are plotted as the blue curves, we then use the MAP estimate to generate the red curve and compare to the data points shown in black. Right: in the off-diagonal panels, we show MCMC scatter plots of pairwise parameter comparisons and, in the diagonal, we show MCMC histograms of each parameter. Our results generate parameter values in line with expectations based on the literature. We also note the ridge like structure of the $(m,a)$ joint distribution indicating a non-identifiability as moves along one domain are highly correlated with the other to maintain a fixed value for $\frac{m}{a}$ in the IVP. }
    \label{fig:real_data}
\end{figure}

Having demonstrated our framework's ability to accurately recover ground truth parameter values with synthetic data, we now demonstrate its application on real laboratory data. Here, our summary statistics $z_{1:N}^{1:M}$ are obtained directly from the batch growth \textit{Pro} data acquired in the experiments of \cref{sec:data_aquisition}. Our data are shown in \cref{fig:real_data} and, as previously, we distinguish a case that considers only batch means (upper panels) and one that considers both batch means and standard deviations (lower panels). For the two cases, our MAP estimators are $(\hat Q,\hat P,\hat m,\hat a) = (113\,000,153,1.114,0.000\,0699)$ and $(\hat Q,\hat P,\hat m,\hat a) = (134\,000,319,0.567,0.000\,0964)$, respectively. These represent growth rate estimates in line with the empirical literature (c.f.~Figure 2 of \cite{martiny2016interactions}, Figure~S5 of \cite{johnson2006niche}).
In addition to MAP estimates, our framework fully quantifies uncertainty in this case in either trajectories or parameter values. As anticipated,
similar to the synthetic cases of \cref{sec:in_silico}, our learning framework also performs well with real experimental data.

\subsection{Sensitivity analysis}

Because our Bayesian framework depends on prior probability distributions, which may influence our posterior estimates, for all analyses shown we choose weakly informative priors for the parameters \cite{gelman1995}. We achieve these using in all simulations default values for the shape hyperparameters $\phi_Q=\phi_P=\phi_m=\phi_a=2$ in \cref{eq:prior_Q,eq:prior_P,eq:prior_m,eq:prior_a}. However, the expectation hyperparameters $\psi_Q,\psi_P,\psi_m,\psi_a$ may still have a substantial effect. For this reason, to quantify our prior's influence on the resulting estimates, we conduct a sensitivity analysis taking into account large changes in their values. 

\begin{figure}[tbp]
    \centering
\includegraphics[width=\textwidth]{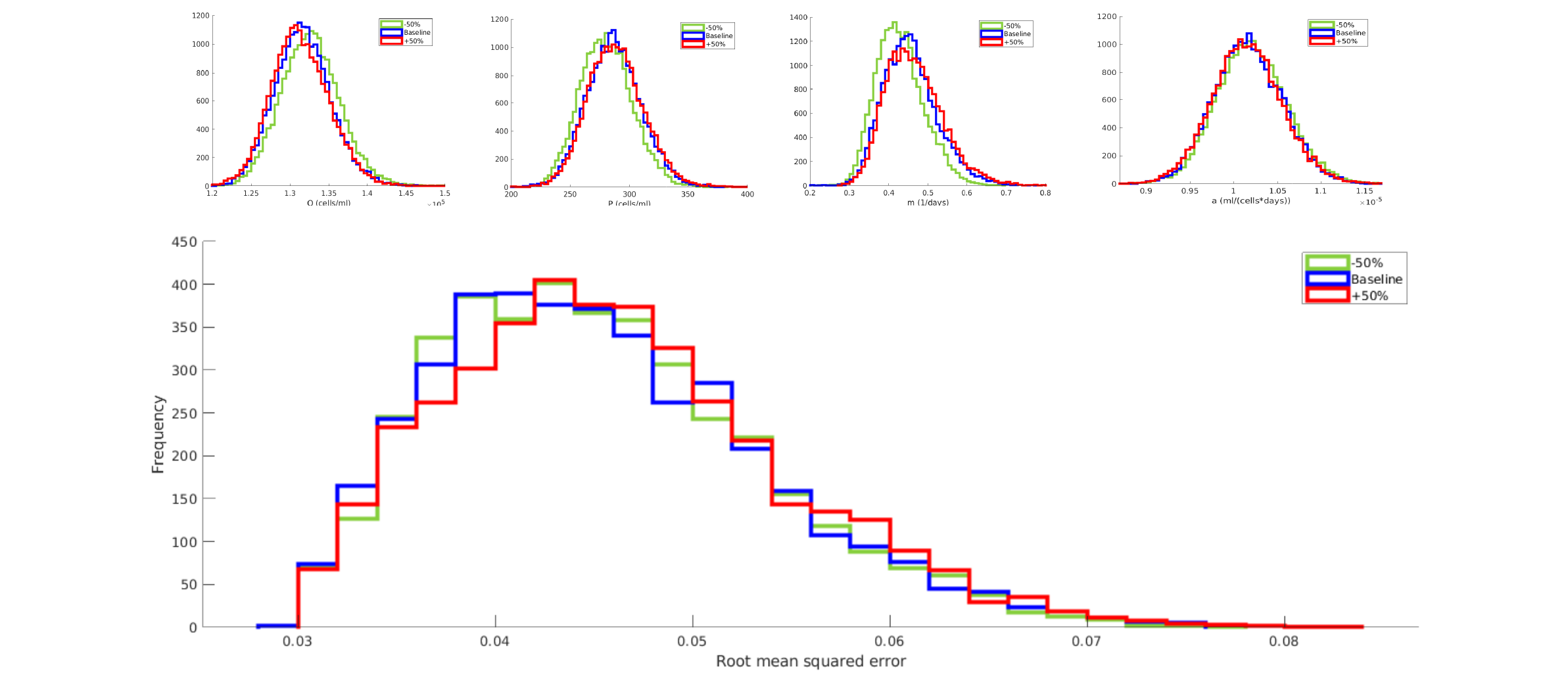}
    \caption{Sensitivity analysis of our posterior estimates of the hyperparameters. To evaluate the sensitivity, we change the values of the hyperparameter $\psi_Q,\psi_P,\psi_m,\psi_a$ from a baseline (blue) to an increase (red) of $50\%$ and a decrease (green) of $50\%$. In the upper panels, we show marginal posteriors for each parameter across the three cases. In the lower panel, we show the root mean squared error (RMSE) when we compare our MCMC trajectories from the three different cases with the data. Across all panels, we see a significant overlap of the three cases, indicating that our methods are insensitive to changes in the hyperparameter values. }
    \label{fig:sensitivity_analysis}
\end{figure}

Specifically, we vary the values of $\psi_Q,\psi_P,\psi_m,\psi_a$ within $\pm50\%$ of their baseline values used in our earlier simulations and derive the resulting marginal posteriors $\mathcal{P}(Q|z_{1:N}^{1:M}),\mathcal{P}(P|z_{1:N}^{1:M}),\mathcal{P}(m|z_{1:N}^{1:M}),\mathcal{P}(a|z_{1:N}^{1:M})$. As shown in \cref{fig:sensitivity_analysis} (top panels), our hyperparameter changes are \emph{not} transmitted to the resulting posteriors, indicating that the associated point estimates are robust, as expected, to the choice of the priors.

In addition, a quantitative comparison of the estimated MAP trajectories $p^{\hat{g}}(\cdot)$ via the relative root-mean-square error
\begin{equation*}
    \text{RMSE}
    = \sqrt{\frac{1}{N}\sum_{n}\left(\frac{p^{\hat g}(t_n)}{z_{n}} -1 \right)^2},
\end{equation*}
also shown on \cref{fig:sensitivity_analysis} (bottom panel), illustrates that hyperparameter changes are not transmitted to the trajectories as well. This indicates that the dynamics recovered by our framework are informed by the supplied data rather than by the hyperparameters.

\section{Discussion}

In this study, we present a unified statistical learning approach to the estimation of the parameters of IVPs probed with aggregated data. Our framework reproduces the missing measurements and explicitly accounts for data aggregation, this way allowing for high modeling fidelity and improved estimation. For our learning tasks, we derived and applied specialized MCMC methods, such as cHMC and mESS, that include a modified HMC sampler to handle summary statistics constraints in the data we model, as well as a modified elliptical slice sampler to navigate parameter sampling \cite{betancourt2017conceptual,neal2012mcmc,murray2010elliptical,cabezas2023transport}. Developed within a fully Bayesian methodology, our framework can be readily used to provide point estimates of the parameters of interest as well as uncertainty quantification.

Our study focuses on the analysis of biological data derived from replication measurements as commonly used in biological research. We demonstrated that our methods successfully analyze data from laboratory batch growth experiments conducted on \textit{Prochlorococcus} cultures and a specialized dynamical system representing the underlying behavior. Nevertheless, the generality of our method allows for applications involving diverse data sets or dynamical models beyond those tested in this study.

Due to its robustness, our methods can contribute to principled learning and assimilation efforts in IVPs where the underlying measurements are hidden, missing, or distorted by non-Gaussian non-additive noise and aggregation. These features are critical for extracting reliable insights from historical data sets or data sets that are confidential or obfuscated, and only down-sampled data or summary statistics are available. Using our advanced learning approach, our methods provide modelers and practitioners with enhanced tools to decipher incomplete data, ensuring the validity of their analysis despite the lack of pre-aggregated information. Our approach not only strengthens the accuracy of the inference that can be drawn, but also facilitates broader applications in the biological and life sciences \cite{taylor2022comparison,jiang2018proper,mante2019heuristic}.

Our novel framework leverages the Bayesian paradigm to unify all aspects of modeling biological dynamics under uncertainty within a single posterior distribution \cite{presse2023}. Our comprehensive approach allows simultaneous parameter estimation, recovery of lost information, and interpretation in physical terms; however, its improved performance comes with its drawbacks. In particular, our framework entails a complex statistical learning approach that requires an extensive mathematical background and advanced computational procedures that pose a barrier to practitioners. To help a wider adoption of the developed methods, we provide a \emph{prototype computational implementation} in \cite{mccoy2025bioivp}. Our implementation is developed in the Matlab programming language \cite{MATLAB} and solves the synthetic \emph{Prochlorococcus} scenarios of \cref{sec:meth_pro,sec:in_silico}.

Nevertheless, a successful implementation of our framework cannot avoid costly algorithmic steps, such as the numerical integration of the underlying ODE, or, similar to all methods based on MCMC, repetitive generation of random variates leading to generally long runs even under the efficient sampling schemes developed herewith. Characteristically, the generation of the results in each analysis of this study with our prototype implementation requires runs of $\approx 5$~min, on a typical single-core laptop computer (Apple MacBook Pro, 2022 model). This time is contrasted with least squares fitting that is completed within $\approx 1$~min. Reducing such a high computational time is the focus of future research that may consider specialized algorithmic solutions within problem-specific domains.

\section*{Conflict of interest}

The authors declare no conflict of interest. 

\section*{Acknowledgments}
This research is based on raw data provided by Erik Zinser through laboratory work supported by the NSF OCE-2023680 grant.
  
\section*{Author contributions}

SM, DM, IS developed statistical methods and wrote the manuscript.
DT and KM commented on the manuscript.
SM contributed to the mathematical formulation and software of mESS.
DM contributed to the mathematical formulation and software of cHMC.
BC and EZ provided the experimental batch culture data.
IS oversaw all aspects of the project.


\bibliographystyle{plain}
\bibliography{ref}

\begin{thebibliography}{10}

\bibitem{almutiry2021continuous}
Waleed Almutiry, Vineetha~Warriyar KV, and Rob Deardon.
\newblock Continuous time individual-level models of infectious disease:
  Package epiilmct.
\newblock {\em Journal of Statistical Software}, 98:1--44, 2021.

\bibitem{A1983}
Hans~C. Andersen.
\newblock Rattle: a "velocity" version of the shake algorithm for molecular
  dynamics calculations.
\newblock {\em Journal of Computational Physics}, 52:24--34, 1983.

\bibitem{atkinson1991}
Kendall Atkinson.
\newblock {\em An introduction to numerical analysis}.
\newblock John wiley \& sons, 1991.

\bibitem{berube2015physiology}
Paul~M Berube, Steven~J Biller, Alyssa~G Kent, Jessie~W Berta-Thompson, Sara~E
  Roggensack, Kathryn~H Roache-Johnson, Marcia Ackerman, Lisa~R Moore, Joshua~D
  Meisel, Daniel Sher, et~al.
\newblock Physiology and evolution of nitrate acquisition in prochlorococcus.
\newblock {\em The ISME journal}, 9(5):1195--1207, 2015.

\bibitem{betancourt2017conceptual}
Michael Betancourt.
\newblock A conceptual introduction to {Hamiltonian Monte Carlo}.
\newblock {\em arXiv preprint arXiv:1701.02434}, 2017.

\bibitem{biller2015}
Steven~J Biller, Paul~M Berube, Debbie Lindell, and Sallie~W Chisholm.
\newblock Prochlorococcus: the structure and function of collective diversity.
\newblock {\em Nature Reviews Microbiology}, 13(1):13--27, 2015.

\bibitem{BN2006}
Christopher~M. Bishop and N.~M. Nasrabdi.
\newblock {\em Pattern Recognition and Machine Learning}.
\newblock Springer, 2006.

\bibitem{briers2010smoothing}
Mark Briers, Arnaud Doucet, and Simon Maskell.
\newblock Smoothing algorithms for state--space models.
\newblock {\em Annals of the Institute of Statistical Mathematics}, 62:61--89,
  2010.

\bibitem{GJM2011}
S.~Brooks, A.~Gelman, and X.~Meng, editors.
\newblock {\em Handbook of Markov Chain Monte Carlo}.
\newblock CRC Press, 2011.

\bibitem{B2012}
Marcus Brubaker, Mathieu Salzmann, and Raquel Urtasun.
\newblock A family of {MCMC} methods on implicitly defined manifolds.
\newblock In {\em Artificial intelligence and statistics}, pages 161--172.
  PMLR, 2012.

\bibitem{cabezas2023transport}
Alberto Cabezas and Christopher Nemeth.
\newblock Transport elliptical slice sampling.
\newblock In {\em International Conference on Artificial Intelligence and
  Statistics}, pages 3664--3676. PMLR, 2023.

\bibitem{calfee2022prochlorococcus}
Benjamin~C Calfee, Liz~D Glasgo, and Erik~R Zinser.
\newblock Prochlorococcus exudate stimulates heterotrophic bacterial
  competition with rival phytoplankton for available nitrogen.
\newblock {\em MBio}, 13(1):e02571--21, 2022.

\bibitem{campbell1995}
Janet~W Campbell.
\newblock The lognormal distribution as a model for bio-optical variability in
  the sea.
\newblock {\em Journal of Geophysical Research: Oceans}, 100(C7):13237--13254,
  1995.

\bibitem{campbell1995lognormal}
Janet~W Campbell.
\newblock The lognormal distribution as a model for bio-optical variability in
  the sea.
\newblock {\em Journal of Geophysical Research: Oceans}, 100(C7):13237--13254,
  1995.

\bibitem{cavender1998dual}
Kent~K Cavender-Bares, Sheila~L Frankel, and Sallie~W Chisholm.
\newblock A dual sheath flow cytometer for shipboard analyses of phytoplankton
  communities from the oligotrophic oceans.
\newblock {\em Limnology and oceanography}, 43(6):1383--1388, 1998.

\bibitem{clark1976effects}
William~AV Clark and Karen~L Avery.
\newblock The effects of data aggregation in statistical analysis.
\newblock {\em Geographical Analysis}, 8(4):428--438, 1976.

\bibitem{diaz2017}
Jes{\'u}s D{\'\i}az~Garc{\'\i}a, Pere Brunet~Crosa, Isabel Navazo~{\'A}lvaro,
  and Pere~Pau V{\'a}zquez~Alcocer.
\newblock Downsampling methods for medical datasets.
\newblock In {\em Proceedings of the International conferences Computer
  Graphics, Visualization, Computer Vision and Image Processing 2017 and Big
  Data Analytics, Data Mining and Computational Intelligence 2017: Lisbon,
  Portugal, July 21-23, 2017}, pages 12--20. IADIS Press, 2017.

\bibitem{djuric2003particle}
Petar~M Djuric, Jayesh~H Kotecha, Jianqui Zhang, Yufei Huang, Tadesse Ghirmai,
  M{\'o}nica~F Bugallo, and Joaquin Miguez.
\newblock Particle filtering.
\newblock {\em IEEE signal processing magazine}, 20(5):19--38, 2003.

\bibitem{flombaum2013present}
Pedro Flombaum, Jos{\'e}~L Gallegos, Rodolfo~A Gordillo, Jos{\'e} Rinc{\'o}n,
  Lina~L Zabala, Nianzhi Jiao, David~M Karl, William~KW Li, Michael~W Lomas,
  Daniele Veneziano, et~al.
\newblock Present and future global distributions of the marine cyanobacteria
  prochlorococcus and synechococcus.
\newblock {\em Proceedings of the National Academy of Sciences},
  110(24):9824--9829, 2013.

\bibitem{gabor2015robust}
Attila G{\'a}bor and Julio~R Banga.
\newblock Robust and efficient parameter estimation in dynamic models of
  biological systems.
\newblock {\em BMC systems biology}, 9:1--25, 2015.

\bibitem{gelman1995}
Andrew Gelman, John~B Carlin, Hal~S Stern, and Donald~B Rubin.
\newblock {\em Bayesian data analysis}.
\newblock Chapman and Hall/CRC, 1995.

\bibitem{GPS2002}
H.~Goldstein, C.~Poole, and J.~Safko.
\newblock {\em Classical Mechanics}.
\newblock Addison-Wesley, 1980.

\bibitem{haibe2020}
Benjamin Haibe-Kains, George~Alexandru Adam, Ahmed Hosny, Farnoosh Khodakarami,
  Massive Analysis Quality Control (MAQC) Society~Board of~Directors Shraddha
  Thakkar 35 Kusko Rebecca 36 Sansone Susanna-Assunta 37 Tong Weida 35
  Wolfinger Russ D. 38 Mason Christopher E. 39 Jones Wendell 40 Dopazo Joaquin
  41 Furlanello Cesare~42, Levi Waldron, Bo~Wang, Chris McIntosh, Anna
  Goldenberg, Anshul Kundaje, et~al.
\newblock Transparency and reproducibility in artificial intelligence.
\newblock {\em Nature}, 586(7829):E14--E16, 2020.

\bibitem{HLW2006}
Ernst Hairer, Christian Lubich, and Gerhard Wanner.
\newblock {\em Geometric Numerical Integration: Structure-Preserving Algorithms
  for Ordinary Differential Equations}.
\newblock Springer, second edition, 2006.

\bibitem{heesche2022implications}
Emil Heesche and Mette Asmild.
\newblock Implications of aggregation uncertainty in data envelopment analysis:
  An application in incentive regulation.
\newblock {\em Decision Analytics Journal}, 4:100103, 2022.

\bibitem{hinson2023}
Audra Hinson, Spiro Papoulis, Lucas Fiet, Margaret Knight, Priscilla Cho,
  Brielle Szeltner, Ioannis Sgouralis, and David Talmy.
\newblock A model of algal-virus population dynamics reveals underlying
  controls on material transfer.
\newblock {\em Limnology and Oceanography}, 68(1):165--180, 2023.

\bibitem{huang2020bayesian}
Hanwen Huang, Andreas Handel, and Xiao Song.
\newblock A bayesian approach to estimate parameters of ordinary differential
  equation.
\newblock {\em Computational statistics}, 35:1481--1499, 2020.

\bibitem{MATLAB}
The~MathWorks Inc.
\newblock Matlab version: 9.13.0 (r2022b), 2022.

\bibitem{jiang2018proper}
Yu~Jiang, Sai Chen, Daniel McGuire, Fang Chen, Mengzhen Liu, William~G Iacono,
  John~K Hewitt, John~E Hokanson, Kenneth Krauter, Markku Laakso, et~al.
\newblock Proper conditional analysis in the presence of missing data:
  Application to large scale meta-analysis of tobacco use phenotypes.
\newblock {\em PLoS genetics}, 14(7):e1007452, 2018.

\bibitem{jim2020}
Heather~SL Jim, Aasha~I Hoogland, Naomi~C Brownstein, Anna Barata, Adam~P
  Dicker, Hans Knoop, Brian~D Gonzalez, Randa Perkins, Dana Rollison, Scott~M
  Gilbert, et~al.
\newblock Innovations in research and clinical care using patient-generated
  health data.
\newblock {\em CA: a cancer journal for clinicians}, 70(3):182--199, 2020.

\bibitem{johnson2006niche}
Zackary~I Johnson, Erik~R Zinser, Allison Coe, Nathan~P McNulty, E~Malcolm~S
  Woodward, and Sallie~W Chisholm.
\newblock Niche partitioning among prochlorococcus ecotypes along ocean-scale
  environmental gradients.
\newblock {\em Science}, 311(5768):1737--1740, 2006.

\bibitem{kilic2021continuous}
Zeliha Kilic, Ioannis Sgouralis, Wooseok Heo, Kunihiko Ishii, Tahei Tahara, and
  Steve Press{\'e}.
\newblock A continuous time representation of smfret for the extraction of
  rapid kinetics.
\newblock {\em Biophysical Journal}, 120(3):186a, 2021.

\bibitem{kilic2021extraction}
Zeliha Kilic, Ioannis Sgouralis, Wooseok Heo, Kunihiko Ishii, Tahei Tahara, and
  Steve Press{\'e}.
\newblock Extraction of rapid kinetics from smfret measurements using
  integrative detectors.
\newblock {\em Cell Reports Physical Science}, 2(5), 2021.

\bibitem{kilic2021generalizing}
Zeliha Kilic, Ioannis Sgouralis, and Steve Press{\'e}.
\newblock Generalizing hmms to continuous time for fast kinetics: Hidden markov
  jump processes.
\newblock {\em Biophysical journal}, 120(3):409--423, 2021.

\bibitem{kilic2021residence}
Zeliha Kilic, Ioannis Sgouralis, and Steve Press{\'e}.
\newblock Residence time analysis of rna polymerase transcription dynamics: A
  bayesian sticky hmm approach.
\newblock {\em Biophysical journal}, 120(9):1665--1679, 2021.

\bibitem{KLSV2022}
Yunbum et~al Kook.
\newblock Sampling with {Riemannian Hamiltonian Monte Carlo} in a constrained
  space.
\newblock {\em Proceedings of the 36th Conference on Neural Processing
  Systems}, 2022.

\bibitem{lagarias1998}
Jeffrey~C Lagarias, James~A Reeds, Margaret~H Wright, and Paul~E Wright.
\newblock Convergence properties of the nelder--mead simplex method in low
  dimensions.
\newblock {\em SIAM Journal on optimization}, 9(1):112--147, 1998.

\bibitem{law2015}
Kody Law, Andrew Stuart, and Kostas Zygalakis.
\newblock Data assimilation.
\newblock {\em Cham, Switzerland: Springer}, 214:52, 2015.

\bibitem{lee2017}
Antony Lee, Konstantinos Tsekouras, Christopher Calderon, Carlos Bustamante,
  and Steve Press{\'e}.
\newblock Unraveling the thousand word picture: an introduction to
  super-resolution data analysis.
\newblock {\em Chemical reviews}, 117(11):7276--7330, 2017.

\bibitem{lee1989}
Peter~M Lee.
\newblock {\em Bayesian statistics}.
\newblock Oxford University Press London:, 1989.

\bibitem{LR1994}
B.~Leimkuhler and S.~Reich.
\newblock Symplectic integration of constrained {Hamiltonian} systems.
\newblock {\em Mathematics of Computation}, 63(208):589--605, October 1994.

\bibitem{LR2004}
Benedict Leimkuhler and Sebastian Reich.
\newblock {\em Simulating Hamiltonian Dynamics}.
\newblock Cambridge University Press, 2004.

\bibitem{LRS2019}
Tony Leli{\'e}vre, Mathias Rousset, and Gabriel Stoltz.
\newblock Hybrid {Monte Carlo} methods for sampling probability measures on
  submanifolds.
\newblock {\em Numerische Mathematik}, 143:379--421, 2019.

\bibitem{linden2022}
Nathaniel~J Linden, Boris Kramer, and Padmini Rangamani.
\newblock Bayesian parameter estimation for dynamical models in systems
  biology.
\newblock {\em PLOS Computational Biology}, 18(10):e1010651, 2022.

\bibitem{L2001}
J.~S. Liu.
\newblock {\em Monte Carlo Strategies in Scientific Computing}.
\newblock Springer, 2001.

\bibitem{maiwald2016driving}
Tim Maiwald, Helge Hass, Bernhard Steiert, Joep Vanlier, Raphael Engesser,
  Andreas Raue, Friederike Kipkeew, Hans~H Bock, Daniel Kaschek, Clemens
  Kreutz, et~al.
\newblock Driving the model to its limit: profile likelihood based model
  reduction.
\newblock {\em PloS one}, 11(9):e0162366, 2016.

\bibitem{mante2019heuristic}
Jeanet Mante, Nishanthi Gangadharan, David~J Sewell, Richard Turner, Ray Field,
  Stephen~G Oliver, Nigel Slater, and Duygu Dikicioglu.
\newblock A heuristic approach to handling missing data in biologics
  manufacturing databases.
\newblock {\em Bioprocess and biosystems engineering}, 42:657--663, 2019.

\bibitem{markham2023}
Katherine Markham, Amy~E Frazier, Kunwar~K Singh, and Marguerite Madden.
\newblock A review of methods for scaling remotely sensed data for spatial
  pattern analysis.
\newblock {\em Landscape Ecology}, 38(3):619--635, 2023.

\bibitem{martiny2016interactions}
Adam~C Martiny, Lanying Ma, C{\'e}line Mouginot, Jeremy~W Chandler, and Erik~R
  Zinser.
\newblock Interactions between thermal acclimation, growth rate, and phylogeny
  influence prochlorococcus elemental stoichiometry.
\newblock {\em PLoS One}, 11(12):e0168291, 2016.

\bibitem{mattamira2025bayesian}
Chiara Mattamira, Alyssa Ward, Sriram~Tiruvadi Krishnan, Rajan Lamichhane,
  Francisco~N Barrera, and Ioannis Sgouralis.
\newblock Bayesian analysis and efficient algorithms for single-molecule
  fluorescence data and step counting.
\newblock {\em bioRxiv}, pages 2025--03, 2025.

\bibitem{mccoy2025bioivp}
Stephen McCoy, Daniel McBride, and Ioannis Sgouralis.
\newblock bio\_ivp\_mcmc repository.
\newblock \url{https://github.com/sgouralis-research-group/bio_IVP_MCMC}, 2025.
\newblock GitHub repository.

\bibitem{metropolis1953}
Nicholas Metropolis, Arianna~W Rosenbluth, Marshall~N Rosenbluth, Augusta~H
  Teller, and Edward Teller.
\newblock Equation of state calculations by fast computing machines.
\newblock {\em The journal of chemical physics}, 21(6):1087--1092, 1953.

\bibitem{moore2007culturing}
Lisa~R Moore, Allison Coe, Erik~R Zinser, Mak~A Saito, Matthew~B Sullivan,
  Debbie Lindell, Katya Frois-Moniz, John Waterbury, and Sallie~W Chisholm.
\newblock Culturing the marine cyanobacterium prochlorococcus.
\newblock {\em Limnology and Oceanography: Methods}, 5(10):353--362, 2007.

\bibitem{morris2011dependence}
J~Jeffrey Morris, Zackary~I Johnson, Martin~J Szul, Martin Keller, and Erik~R
  Zinser.
\newblock Dependence of the cyanobacterium prochlorococcus on hydrogen peroxide
  scavenging microbes for growth at the ocean's surface.
\newblock {\em PloS one}, 6(2):e16805, 2011.

\bibitem{morris2008facilitation}
J~Jeffrey Morris, Robin Kirkegaard, Martin~J Szul, Zackary~I Johnson, and
  Erik~R Zinser.
\newblock Facilitation of robust growth of prochlorococcus colonies and dilute
  liquid cultures by “helper” heterotrophic bacteria.
\newblock {\em Applied and environmental microbiology}, 74(14):4530--4534,
  2008.

\bibitem{murphy2024implementing}
Ryan~J Murphy, Oliver~J Maclaren, and Matthew~J Simpson.
\newblock Implementing measurement error models with mechanistic mathematical
  models in a likelihood-based framework for estimation, identifiability
  analysis and prediction in the life sciences.
\newblock {\em Journal of the Royal Society Interface}, 21(210):20230402, 2024.

\bibitem{murray2010elliptical}
Iain Murray, Ryan Adams, and David MacKay.
\newblock Elliptical slice sampling.
\newblock In {\em Proceedings of the thirteenth international conference on
  artificial intelligence and statistics}, pages 541--548. JMLR Workshop and
  Conference Proceedings, 2010.

\bibitem{neal2012mcmc}
Radford~M Neal.
\newblock Mcmc using hamiltonian dynamics.
\newblock {\em arXiv preprint arXiv:1206.1901}, 2012.

\bibitem{orcutt1968}
Guy~H Orcutt, Harold~W Watts, and John~B Edwards.
\newblock Data aggregation and information loss.
\newblock {\em The American Economic Review}, 58(4):773--787, 1968.

\bibitem{partensky1999proc}
Fred Partensky, Wolfgang~R Hess, and Daniel Vaulot.
\newblock Prochlorococcus, a marine photosynthetic prokaryote of global
  significance.
\newblock {\em Microbiology and molecular biology reviews}, 63(1):106--127,
  1999.

\bibitem{presse2023}
Steve Press{\'e} and Ioannis Sgouralis.
\newblock {\em Data Modeling for the Sciences: Applications, Basics,
  Computations}.
\newblock Cambridge University Press, 2023.

\bibitem{quarteroni2006}
Alfio Quarteroni, Riccardo Sacco, and Fausto Saleri.
\newblock {\em Numerical mathematics}, volume~37.
\newblock Springer Science \& Business Media, 2006.

\bibitem{raue2013lessons}
Andreas Raue, Marcel Schilling, Julie Bachmann, Andrew Matteson, Max Schelke,
  Daniel Kaschek, Sabine Hug, Clemens Kreutz, Brian~D Harms, Fabian~J Theis,
  et~al.
\newblock Lessons learned from quantitative dynamical modeling in systems
  biology.
\newblock {\em PloS one}, 8(9):e74335, 2013.

\bibitem{reich2015probabilistic}
Sebastian Reich and Colin Cotter.
\newblock {\em Probabilistic forecasting and Bayesian data assimilation}.
\newblock Cambridge University Press, 2015.

\bibitem{RC1999}
Christian~P. Robert and G.~Casella.
\newblock {\em Monte Carlo Statistical Methods}.
\newblock Springer, 1999.

\bibitem{roda2020bayesian}
Weston~C Roda.
\newblock Bayesian inference for dynamical systems.
\newblock {\em Infectious Disease Modelling}, 5:221--232, 2020.

\bibitem{rodriguez2006hybrid}
Maria Rodriguez-Fernandez, Pedro Mendes, and Julio~R Banga.
\newblock A hybrid approach for efficient and robust parameter estimation in
  biochemical pathways.
\newblock {\em Biosystems}, 83(2-3):248--265, 2006.

\bibitem{ronan2016}
Tom Ronan, Zhijie Qi, and Kristen~M Naegle.
\newblock Avoiding common pitfalls when clustering biological data.
\newblock {\em Science signaling}, 9(432):re6--re6, 2016.

\bibitem{salgado2022classical}
Abner~J Salgado and Steven~M Wise.
\newblock {\em Classical numerical analysis: a comprehensive course}.
\newblock Cambridge University Press, 2022.

\bibitem{schober2014probabilistic}
Michael Schober, David~K Duvenaud, and Philipp Hennig.
\newblock Probabilistic ode solvers with runge-kutta means.
\newblock {\em Advances in neural information processing systems}, 27, 2014.

\bibitem{sgouralis2017renal}
Ioannis Sgouralis, Roger~G Evans, and Anita~T Layton.
\newblock Renal medullary and urinary oxygen tension during cardiopulmonary
  bypass in the rat.
\newblock {\em Mathematical medicine and biology: a journal of the IMA},
  34(3):313--333, 2017.

\bibitem{sgouralis2016bladder}
Ioannis Sgouralis, Michelle~M Kett, Connie~PC Ow, Amany Abdelkader, Anita~T
  Layton, Bruce~S Gardiner, David~W Smith, Yugeesh~R Lankadeva, and Roger~G
  Evans.
\newblock Bladder urine oxygen tension for assessing renal medullary
  oxygenation in rabbits: experimental and modeling studies.
\newblock {\em American Journal of Physiology-Regulatory, Integrative and
  Comparative Physiology}, 311(3):R532--R544, 2016.

\bibitem{sgouralis2013control}
Ioannis Sgouralis and Anita~T Layton.
\newblock Control and modulation of fluid flow in the rat kidney.
\newblock {\em Bulletin of mathematical biology}, 75:2551--2574, 2013.

\bibitem{sgouralis2014theoretical}
Ioannis Sgouralis and Anita~T Layton.
\newblock Theoretical assessment of renal autoregulatory mechanisms.
\newblock {\em American Journal of Physiology-Renal Physiology},
  306(11):F1357--F1371, 2014.

\bibitem{sgouralis2015mathematical}
Ioannis Sgouralis and Anita~T Layton.
\newblock Mathematical modeling of renal hemodynamics in physiology and
  pathophysiology.
\newblock {\em Mathematical biosciences}, 264:8--20, 2015.

\bibitem{sgouralis2016transfer}
Ioannis Sgouralis, Vasileios Maroulas, and Anita~T Layton.
\newblock Transfer function analysis of dynamic blood flow control in the rat
  kidney.
\newblock {\em Bulletin of mathematical biology}, 78(5):923--960, 2016.

\bibitem{sgouralis2017introduction}
Ioannis Sgouralis and Steve Press{\'e}.
\newblock An introduction to infinite hmms for single-molecule data analysis.
\newblock {\em Biophysical journal}, 112(10):2021--2029, 2017.

\bibitem{shampine1997}
Lawrence~F Shampine and Mark~W Reichelt.
\newblock The matlab ode suite.
\newblock {\em SIAM journal on scientific computing}, 18(1):1--22, 1997.

\bibitem{sivia2006data}
Devinderjit Sivia and John Skilling.
\newblock {\em Data analysis: a Bayesian tutorial}.
\newblock OUP Oxford, 2006.

\bibitem{smith2020}
Vinayak Smith, Densearn Seo, Ritesh Warty, Olivia Payne, Mohamed Salih, Ken~Lee
  Chin, Richard Ofori-Asenso, Sathya Krishnan, Fabricio da~Silva~Costa,
  Beverley Vollenhoven, et~al.
\newblock Maternal and neonatal outcomes associated with covid-19 infection: A
  systematic review.
\newblock {\em Plos one}, 15(6):e0234187, 2020.

\bibitem{stoer1980}
Josef Stoer, Roland Bulirsch, R~Bartels, Walter Gautschi, and Christoph
  Witzgall.
\newblock {\em Introduction to numerical analysis}, volume 1993.
\newblock Springer, 1980.

\bibitem{taylor2022comparison}
Sandra Taylor, Matthew Ponzini, Machelle Wilson, and Kyoungmi Kim.
\newblock Comparison of imputation and imputation-free methods for statistical
  analysis of mass spectrometry data with missing data.
\newblock {\em Briefings in Bioinformatics}, 23(1):bbab353, 2022.

\bibitem{tronarp2019probabilistic}
Filip Tronarp, Hans Kersting, Simo S{\"a}rkk{\"a}, and Philipp Hennig.
\newblock Probabilistic solutions to ordinary differential equations as
  nonlinear bayesian filtering: a new perspective.
\newblock {\em Statistics and Computing}, 29:1297--1315, 2019.

\bibitem{van2011}
A~Franciscus van~der Meer, Marco~AE Marcus, Dani{\"e}l~J Touw, Johannes~H
  Proost, and Cees Neef.
\newblock Optimal sampling strategy development methodology using maximum a
  posteriori bayesian estimation.
\newblock {\em Therapeutic drug monitoring}, 33(2):133--146, 2011.

\bibitem{venkataramani2020}
Atheendar~S Venkataramani, Rourke O’Brien, Gregory~L Whitehorn, and
  Alexander~C Tsai.
\newblock Economic influences on population health in the united states: toward
  policymaking driven by data and evidence.
\newblock {\em PLoS medicine}, 17(9):e1003319, 2020.

\bibitem{wieland2021structural}
Franz-Georg Wieland, Adrian~L Hauber, Marcus Rosenblatt, Christian T{\"o}nsing,
  and Jens Timmer.
\newblock On structural and practical identifiability.
\newblock {\em Current Opinion in Systems Biology}, 25:60--69, 2021.

\bibitem{williamson2002}
Paula~R Williamson, Catrin~Tudur Smith, Jane~L Hutton, and Anthony~G Marson.
\newblock Aggregate data meta-analysis with time-to-event outcomes.
\newblock {\em Statistics in medicine}, 21(22):3337--3351, 2002.

\bibitem{wilson2021}
Samantha~L Wilson, Gregory~P Way, Wout Bittremieux, Jean-Paul Armache,
  Melissa~A Haendel, and Michael~M Hoffman.
\newblock Sharing biological data: why, when, and how.
\newblock {\em FEBS letters}, 595(7):847, 2021.

\bibitem{xia2022}
Fangfang Xia, Jonathan Allen, Prasanna Balaprakash, Thomas Brettin, Cristina
  Garcia-Cardona, Austin Clyde, Judith Cohn, James Doroshow, Xiaotian Duan,
  Veronika Dubinkina, et~al.
\newblock A cross-study analysis of drug response prediction in cancer cell
  lines.
\newblock {\em Briefings in bioinformatics}, 23(1):bbab356, 2022.

\bibitem{xu2020}
Xiangyu Xu, Muchen Li, Wenxiu Sun, and Ming-Hsuan Yang.
\newblock Learning spatial and spatio-temporal pixel aggregations for image and
  video denoising.
\newblock {\em IEEE Transactions on Image Processing}, 29:7153--7165, 2020.

\bibitem{zinser2009choreography}
Erik~R Zinser, Debbie Lindell, Zackary~I Johnson, Matthias~E Futschik, Claudia
  Steglich, Maureen~L Coleman, Matthew~A Wright, Trent Rector, Robert Steen,
  Nathan McNulty, et~al.
\newblock Choreography of the transcriptome, photophysiology, and cell cycle of
  a minimal photoautotroph, prochlorococcus.
\newblock {\em PloS one}, 4(4):e5135, 2009.

\end{thebibliography}

\newpage
\appendix
\counterwithin{figure}{section}
\renewcommand\thefigure{S.\arabic{figure}} 
\counterwithin{table}{section}
\renewcommand\thetable{S.\arabic{table}}

\section{Summary of the framework's equations}
\label{sec:append_eqn}

In its most general form, our framework is described by:
\begin{align}
    g &\sim \mathbb{C} \\
    h &\sim \mathbb{B} \\
    y_{n}^{k}|g,h &\sim \mathbb{A}\left(F\left(x^g(t_n)\right),h\right)  & k=1,\dots,K,\quad n=1,\dots,N  \\
    z_n\mid y_n^{1:K} &\sim \d_{G(y_n^{1:K})} & n=1,\dots,N
\end{align}
where $\mathbb{B},\mathbb{C}$ are the prior distributions of the dynamical $g$ and observation $h$ parameters, $\mathbb{A}$ is the likelihood of the raw measurements $y_{1:N}^{1:M}$, $F(\cdot)$ is the observation function, and $x^g(\cdot)$ is the solution of the underlying IVP with initial conditions applied at time $t_0$. In this description, the aggregated data is $z_{1:N}^{1:M}$, the evaluation times are $t_{1:N}$, the statistics are $G^{1:M}(\cdot)$, and the batch size is $K$.

\section{Definitions of probability distributions}
\label{sec:dist_def}

Here we present the parameterization of the probability distributions that we use throughout this study.

\textbullet~The \emph{Log Normal} has probability density given by
\begin{equation*}
    \LogN(y;x,h) = \frac{1}{y}\sqrt{\frac{h}{2\pi}}\exp\left(-\frac{h}{2}\left(\log\frac{y}{x}\right)^2\right)
\end{equation*}
with $x$ being the median and $h$ being the precision.
\\
\textbullet~The \emph{Gamma} has probability density given by
\begin{equation*}
    \Gam(\g;\k,\th) = \frac{\g^{\k-1}e^{-\g/\th}}{\Gamma(\k)\th^\k}
\end{equation*}
this has mean $\kappa \theta$.
\\
\textbullet~The \emph{Delta} density is given by 
\begin{align*}
    \delta_{a}(E) =
    \begin{cases}
        1 & \text{if } a = E \\
        0 & \text{if } a \neq E
    \end{cases}.
\end{align*}
\\
\textbullet~ The \emph{Bernoulli} probability mass function is
\begin{align*}
    \text{Bernoulli}(k;p) = \begin{cases}
        p & \text{if } k=1, \\ 1-p & \text{if } k=0 \\ 0 & \text{otherwise}
    \end{cases}.
\end{align*}
\\
\textbullet~ The \emph{Uniform} probability density on the interval $(a,b)$ is
\begin{align*}
    \text{Uniform}_{(a,b)}(x) = \begin{cases}
        \frac{1}{b-a} &    \text{if } x\in(a,b) \\
        0 & \text{otherwise}
    \end{cases}.
\end{align*}
\\
\textbullet~ The $d$-dimensional \emph{Normal} probability density is
\begin{align*}
    \text{Normal}(x;\mu,\Sigma) = \frac{1}{\sqrt{(2\pi)^d|\Sigma|}}\exp\left(-\frac{1}{2}(x-\mu)^T\Sigma^{-1}(x-\mu)\right).
\end{align*}

\section{Computational methods}
\label{sec:append_comp}

\subsection{Multiplicative elliptical slice sampling}
\label{sec:append_MESS}

To sample from a generic target with probability density $\pi(g_{1:M})$ over a continuous multivariate random variable $g_{1:M}$, we start by completing with normal auxiliary variables $\lambda_{1:M}$, which are independent from our parameters
\begin{align*}
    \lambda_{m} &\sim \text{Normal}(0,\sigma^2_m), &
    m&=1,\ldots,M.
\end{align*}
After this completion, our target becomes $p(g_{1:M},\lambda_{1:M})$. At this stage, we continue with a 2-step Gibbs sampler consisting of:
\\
\textbullet~Obtain $\lambda_{1:M}$ by sampling from $p(\lambda_{1:M}|g_{1:M})$. Due to the independence between $\lambda_{1:M}$ and $g_{1:M}$, this step is achieved by direct sampling normal variates.
\\
\textbullet~Apply a change of variables $(g_{1:M},\lambda_{1:M})\mapsto(\phi_{1:M},\psi_{1:M})$ defined by
\begin{align*}
    \phi_{m} &= g_{m} \exp(\psi_m),
\\
    \psi_m &= \lambda_m.
\end{align*}
to obtain a transformed target. According to \cite{presse2023}, this is given by
\begin{align*}
    p(\phi_{1:M},\psi_{1:M})
   \propto  \pi(g_{1:M})\exp\left(-\sum_{m=1}^M \lambda_m  \right) 
   \prod_{m=1}^M\text{Normal}(\lambda_m;0,\sigma^2_m)
   .
\end{align*}
\textbullet~Resample $\psi_{1:M}$ from $p(\psi_{1:M}|\phi_{1:M})$. Due to the latent Gaussian form of the transformed target $p(\phi_{1:M},\psi_{1:M})$, this is achieved using the elliptical slice sampler \cite{murray2010elliptical,cabezas2023transport,neal2012mcmc}.
\\
\textbullet~Recover the original variables $g_{1:M}$ using the inverse transform $(\phi_{1:M},\psi_{1:M})\mapsto(g_{1:M},\lambda_{1:M})$ given by
\begin{align*}
    g_{m}&=\phi_{m}\exp(-\psi_m).
\end{align*}
Our entire scheme, including the elliptical slice sampling stage, is implemented in \cref{alg:MESS}.

\begin{algorithm}
\caption{Multiplicative Elliptical Slice Sampler}
\label{alg:MESS}
\begin{algorithmic}
\State \textbf{Input:} Current state $\mathbf{g}$,
auxiliary variance $\sigma^2$, likelihood function $L(\cdot)$
\State \textbf{Output:} New state $\mathbf{g}'$
\State $\mathbf{\nu} \sim \text{Normal}(0, S)$ \Comment{Choose ellipse}
\State $\mathbf{\lambda} \sim \text{Normal}(0, \sigma^2)$ \Comment{Choose auxiliary variable}
\State $u \sim \text{Uniform}(0, 1)$
\State $\log y \leftarrow \log L(\mathbf{g}) + \log u$ \Comment{Set log-likelihood threshold}
\State $\theta \sim \text{Uniform}(0, 2\pi)$ \Comment{Draw an initial proposal}
\State $[\theta_{\min}, \theta_{\max}] \leftarrow [\theta - 2\pi, \theta]$ \Comment{Define a bracket}
\While{true}
    \State $\mathbf{\lambda}' \leftarrow \mathbf{\lambda} \cos \theta + \mathbf{\nu} \sin \theta $
    \Comment{Traverse auxiliary space with ellipse}
    \State $\mathbf{g'} \leftarrow \mathbf{g} * \exp\left(\mathbf{\lambda}- \mathbf{\lambda'}\right)$
    \Comment{Reverse CoV}
    \If{$\log L(\mathbf{g'}) + \sum 
    \mathbf{\lambda}> \log y + \sum\mathbf{\lambda'}$} \Comment{Accept}
        \State \textbf{return} $\mathbf{g'}$
    \Else
        \If{$\theta < 0$}
            \State $\theta_{\min} \leftarrow \theta$
        \Else
            \State $\theta_{\max} \leftarrow \theta$
        \EndIf
        \State $\theta \sim \text{Uniform}(\theta_{\min}, \theta_{\max})$ \Comment{Shrink the bracket and try a new point}
    \EndIf
\EndWhile
\end{algorithmic}
\end{algorithm}

\subsection{Standard Hamiltonian Monte Carlo}
\label{sec:append_HMC}

The HMC algorithm is an MCMC method for sampling a target distribution with density \( \pi(q)\propto f(q) \) known only up to a multiplicative constant. We assume that the support \( Q \) of \( f(q) \) is a Euclidean space of fixed dimension. HMC is an extension of the Metropolis algorithm, but with a proposal rule constructed in analogy with classical Hamiltonian mechanics, the dynamics of which are energy conserving, reversible, and measure preserving \cite{HLW2006, GPS2002}. Here, the sequence of samples \( \{q^{(j)}\}_{j=0:J} \) produced by HMC is viewed as a sequence of positions of a particle, or of an ensemble of particles, governed by Hamiltonian dynamics. This analogy introduces auxiliary ``momentum" variables \( \{p^{(j)}\}_{j=0:J} \) and a ``total energy" Hamiltonian function
\begin{align}
H(q,p)=V(q)+K(p),
\end{align}
where we define, by analogy, a ``potential energy" function
\begin{align}
    V(q)=-\log f(q),
\end{align}
and a ``kinetic energy" function
\begin{align}
    K(p)=\half p^T M^{-1} p,
\end{align}
where \( M \) is a ``mass" matrix. This choice for the Hamiltonian function \cite{betancourt2017conceptual, GJM2011} leverages the known gradient information of our target distribution \( \pi(q) \). We can convert our proportionality relationship
\begin{align}
    \pi(q)\propto f(q)
\end{align}
into an equality involving the gradient of our target distribution,
\begin{align}
    \grad \log f(q) = -\grad V(q).
\end{align}
So the method is, given a sample \( q^{(j)} \), generate a proposal sample by evolving dynamics on phase space (\( (q,p)- \)space) governed by Hamilton's equations, which are, in this case,
\begin{align}
    \dot q &= M^{-1}p, & q(0) &= q^{(j)}, \\
    \dot p &= \grad \log f(q),&  p(0) &= p^{(j)}.
\end{align}
We integrate these dynamics for a fixed time, either exactly or approximately, and denote by
\begin{align}
    \left(\tilde q^{(j+1)},\tilde p^{(j+1)}\right) = \Phi\left( q^{(j)}, p^{(j)}\right)
\end{align}
the result of the integration. Here, \( q^{(j)} \) is carried over from the previous sampling step and \( p^{(j)} \) follows the sampling rule
\begin{align}
    p^{(j)}\sim \text{Normal}(0,M),
\end{align}
where $\text{Normal}(0,M)$ is a multivariate normal. For the Metropolis step, we compute the acceptance ratio \( R^{(j)} \) using the expression
\begin{align}
    R^{(j)}=\exp\left(H\left(q^{(j)}, p^{(j)}\right)-H\left(\tilde q^{(j+1)},\tilde p^{(j+1)}\right)\right).
\end{align}
As in the Metropolis algorithm, we accept the proposed sample and set \( q^{(j+1)} = \tilde q^{(j+1)}  \) with probability \( \min(1,R^{(j)}) \), otherwise we set \( q^{(j+1)} = q^{(j)} \). We summarize the entire scheme in \cref{alg:HMC}.

\begin{algorithm}
\caption{Hamiltonian Monte Carlo}\label{alg:HMC}
\begin{algorithmic}
\State \textbf{Input: } $q^{(0)}\in Q$
\State \textbf{Output: } $\{q^{(j)}\}_{j=0:J}$
\For{$j=0:J-1$}
\State $p^{(j)}\sim\text{Normal}(0,M)$
\State $\left(\tilde q^{(j+1)},\tilde p^{(j+1)}\right) \gets \Phi\left( q^{(j)}, p^{(j)}\right)$
\State $\gamma \sim \text{Bernoulli}(\min(1,R^{(j)}))$
\State $q^{(j+1)} \gets \gamma \tilde q^{(j+1)} + (1-\gamma) q^{(j)}$
\EndFor
\end{algorithmic}
\end{algorithm}

Provided the integration is exact or a reversible, measure-preserving  numerical integrator (such as St{\"o}rmer-Verlet) is used, then this scheme satisfies detailed balance \cite{betancourt2017conceptual}, so the statistics of the resulting Markov chain $\{q^{(j)}\}_{j=0:J}$ approach those of the target distribution with density $\pi(q)$.

In practice the exact integration of the Hamiltonian is not possible; nevertheless, better approximate integration of the Hamiltonian leads to higher acceptance rates \cite{GJM2011,betancourt2017conceptual}. The standard numerical integrator for HMC is the explicit, second order St{\"o}rmer-Verlet numerical scheme \cite{L2001, GJM2011, betancourt2017conceptual}. As it is the base algorithm from which RATTLE is derived, we give one step (step size \( h \)) of the St{\"o}rmer-Verlet numerical scheme in \cref{alg:SV}.

\begin{algorithm}
\caption{St{\"o}rmer-Verlet numerical integration (one step, step size $h$)}\label{alg:SV}
\begin{algorithmic}
\State \textbf{Input: } $(q,p) \text{ in phase space}$
\State \textbf{Output: } $(\tilde q,\tilde p)$
\State $\bar p \gets p - \frac{h}{2}\grad V(q)$
\State $\tilde q \gets q + h\grad K(\bar p)$
\State $\tilde p \gets \bar p - \frac{h}{2}\grad V(\tilde q)$
\end{algorithmic}
\end{algorithm}
The St{\"o}rmer-Verlet scheme is indeed reversible and preserves the phase-space measure \cite{GJM2011,betancourt2017conceptual}, so the resulting Markov chain has the desired approximation properties.

\subsection{Constrained Hamiltonian Monte Carlo}
\label{sec:append_cHMC}

In this section, we follow the notation used in \cref{sec:append_HMC} and outline the constrained Hamiltonian Monte Carlo algorithm as described by Brubaker et al. \cite{B2012}. The goal of cHMC is to sample a surface measure, in particular a probability distribution with density \( \pi(q)\propto f(q) \) known only up to a multiplicative constant supported on a differentiable manifold \( \HM \). We take \( \HM \) to be embedded in Euclidean space and defined as the zero set of a constraint function \( q\mapsto c(q)\in\R^d \) with full-rank Jacobian \( J_{c(q)} \) for all \( q\in\HM \). We will use the notation \( \grad c(q) = J_{c(q)}^T \) in analogy to the gradient for the transpose of the Jacobian. As in the unconstrained setting, we introduce a virtual momentum variable \( p \).

To represent data aggregation, the set of values $q=y_{1:N}^{1:K}$ which simultaneously satisfy \cref{eq:stats} is parametrized as the solution set of the following $2N$ constraint functions, $G^1(y_n^{1:K}) = z_n^1$ and $G^2(y_n^{1:K}) = z_n^2$ for $n=1,\dots,N$, where $z_n^1$ and $z_n^2$ are the mean and standard deviation, respectively, of the $K$ measurements taken at time $t_n$. Here, $G^1(\cdot)$ and $G^2(\cdot)$ are as defined in \cref{G1func,G2func}.

It is convenient for the implementation to reparametrize these $2N$ constraints as follows. We constrain samples $q=y_{1:N}^{1:K}$ to lie on the zero locus of
\begin{align}
    c_n^1(q) &= \sum_{k=1}^K y_n^k-K\tilde z_n^1, & n&=1,\dots,N,
    \\
    c_n^2(q) &= \frac{\tilde z_n^1}{2}\left(\frac{1}{\tilde z_n^2}\sum_{k=1}^K (y_n^k)^2-K\right) & n&=1,\dots,N,
\end{align}
where the new constants $\tilde z_n^1$ and $\tilde z_n^2$ are given in terms of $z_n^1$ and $z_n^2$ according to
\begin{align*}
\tilde z_n^1 &= z_n^1
&
n&=1,\dots,N,
\\
\tilde z_n^2 &= (z_n^1)^2+\frac{K-1}{K}(z_n^2)^2
&
n&=1,\dots,N.
\end{align*}
The new constants are the batch raw (uncentered) first and second moments \cite{presse2023} of the measurements recorded at time $t_n$.

For convenience of the notation, we collect all constraints into a column vector-valued function  $c = (c_1^1,\cdots,c_N^1,\allowbreak c_1^2,\cdots,c_N^2)^T$
which reduces the satisfaction of the $2N$ mean and standard deviation constraints to the simple expression $c(q)=0$.

With this formalism, the method is defined by the following update rule. Given a sample \( q^{(j)}\in\HM \), generate a proposal by evolving the differential algebraic equations
\begin{align}
    \dot q &= M^{-1} p, & q(0)&=q^{(j)}, \\
    \dot p &= \grad\log f(q)-\grad c(q)\l(q,p), & p(0) &= p^{(j)}, \\
    & \l(q,p)\in\R^d\text{ s.t. }c(q) = 0.
\end{align}
Here, \( \l \) is a Lagrange multiplier used to enforce that constraint \( c(q)=0 \) is satisfied throughout the trajectory. As before, \( q^{(j)} \) is inherited from the previous sampling step, but now \( p^{(j)} \) is a random vector sampled from a distribution depending on the geometry of the constraint manifold $\HM$.

Let \( T_q\HM \) denote the plane tangent to \( \HM \) at \( q \), and let \( \HN^{(j)} \) denote the surface measure in \( T_{q^{(j)}}\HM \) induced by the normal distribution \( \text{Normal}(0,M) \) in the ambient Euclidean space. We take the initial momentum sample \( p^{(j)} \) from the singular distribution \( \HN^{(j)} \). Under these initial conditions \cite{GPS2002,LR1994,LR2004}, the solution \( (q,p) \) remains in the tangent bundle \( T\HM=\sqcup_{q\in \HM}T_{q}\HM \) of \( \HM \).

We integrate these dynamics for a fixed time and denote the result by
\begin{align}
    \left(\tilde q^{(j+1)},\tilde p^{(j+1)}\right) = \Psi\left( q^{(j)}, p^{(j)}\right).
\end{align}
The integration can be exact or approximate. We compute the Metropolis acceptance ratio \( R^{(j)} \) using the same expression as in the unconstrained case. Our entire scheme is summarized in \cref{alg:cHMC}.

\begin{algorithm}
\caption{Constrained Hamiltonian Monte Carlo}\label{alg:cHMC}
\begin{algorithmic}
\State \textbf{Input: } $q^{(0)}\in \HM$
\State \textbf{Output: } $\{q^{(j)}\}_{j=0:J}$
\For{$j=0:J-1$}
\State $p^{(j)}\sim\HN^{(j)}$
\State $\left(\tilde q^{(j+1)},\tilde p^{(j+1)}\right) \gets \Psi\left( q^{(j)}, p^{(j)}\right)$
\State $\gamma \sim \text{Bernoulli}(\min(1,R^{(j)}))$
\State $q^{(j+1)} \gets \gamma \tilde q^{(j+1)} + (1-\gamma) q^{(j)}$
\EndFor
\end{algorithmic}
\end{algorithm}

Since a numerical integrator with energy-conserving properties that respects the constraint structure is desired, the standard choice for a numerical integrator is the second-order method RATTLE \cite{B2012,LRS2019,KLSV2022}. It has been shown that RATTLE is reversible and preserves Lebesgue surface measure in \( T\HM \), for instance see \cite{LR1994,LR2004}, and hence satisfies detailed balance. So, the Markov chain generated by cHMC has the desired statistical approximation properties. In \cref{alg:RATTLE}, we describe one step of RATTLE with step size \( h \).

\begin{algorithm}
\caption{RATTLE numerical integration (one step, step size $h$)}\label{alg:RATTLE}
\begin{algorithmic}
\State \textbf{Input: } $(q,p) \in T\HM$
\State \textbf{Output: } $(\tilde q,\tilde p)$
\State $\bar p \gets p - \frac{h}{2}[\grad V(q)+\grad c(q)\l_{(q)}],
\quad\l_{(q)}\text{ s.t. } c(q+h\grad K(\bar p)) = 0$
\State $\tilde q \gets q + h\grad K(\bar p)$
\State $\tilde p \gets \bar p - \frac{h}{2}[\grad V(\tilde q)+\grad c(\tilde q)\l_{(p)}], \quad \l_{(p)}\text{ s.t. } \grad K(\tilde p)\cdot \grad c(\tilde q) = 0$
\end{algorithmic}
\end{algorithm}
Though based on St{\"o}rmer-Verlet, the RATTLE scheme is quite computationally distinct. Each step of RATTLE requires, in general, the solving of a nonlinear and a linear system to find the appropriate multipliers \( \l_{(q)} \) and \( \l_{(p)} \), respectively. We implement these solvers with the Newton-Raphson method \cite{salgado2022classical}. The condition on \( \l_{(q)} \) keeps \( \tilde q\in \HM \) while \( \l_{(p)} \) keeps \( \tilde p\in T_q\HM \) ensuring that the output \( (\tilde q,\tilde p) \) remains in \( T\HM \) as desired.

\section{Least squares parameter estimation}
\label{sec:append_LS}

We formulate least-squares (LS) estimation in \cref{sec:meth_pro} as an optimization problem
\begin{align*}
g^* = \text{argmin}_g L(g),
\end{align*}
where the objective function is given by
\begin{equation*}
    L(g) = \sum_{n=1}^{N} \left(
    z_{n} - F\left(x^g(t_n)\right)
    \right)^2
    .
\end{equation*}
Here, $x^g(t_n)$ denotes the solution of the IVP in \cref{eq:ODE} with parameters $g$ evaluated at time $t_n$, $F(\cdot)$ is the observation function in \cref{eq:like}, and $z_n$ denotes the mean summary statistic computed in \cref{eq:stats}. 

For the numerical solution of the minimization problem, we apply a derivative-free method as implemented in the Nelder-Mead simplex algorithm \cite{lagarias1998}.

\end{document}